\RequirePackage{fix-cm}
\documentclass[smallextended]{svjour3}       
\usepackage{srcltx}
\usepackage{authblk} 
\smartqed  
\usepackage{graphicx}
\usepackage{color}
\usepackage{natbib}
\journalname{Experimental Astronomy}
\newcommand{\gq}{{\it \textbf {GrailQuest}}}
\def\ls{{_<\atop^{\sim}}}

\begin{document}

\title{ {\it \textbf{GrailQuest}}: hunting for Atoms of Space and Time hidden in the wrinkle of Space--Time
}
\subtitle{A swarm of nano/micro/small--satellites to probe the ultimate structure of Space--Time and to provide an all--sky monitor to study high--energy astrophysics phenomena}

\titlerunning{{\it \textbf{GrailQuest}}: hunting for Atoms of Space and Time}       


\author[,1,26,27]{L.~Burderi\footnote{burderi@dsf.unica.it}}
\author[,1]{A.~Sanna\footnote{andrea.sanna@dsf.unica.it}}
\author[2,26,27]{T.~Di Salvo}
\author[3]{L.~Amati}
\author[4]{G.~Amelino-Camelia}
\author[5]{M.~Branchesi}
\author[6]{S.~Capozziello}
\author[5]{E.~Coccia}
\author[7]{M.~Colpi}
\author[8]{E.~Costa}
\author[1,26]{N.~D'Amico}
\author[9]{P.~De Bernardis}
\author[6]{M.~De Laurentis}
\author[10]{M.~Della Valle}
\author[11]{H.~Falcke}
\author[12]{M.~Feroci}
\author[13]{F.~Fiore}
\author[14]{F.~Frontera}
\author[2]{A.~F. Gambino}
\author[15]{G.~Ghisellini}
\author[16]{K.~C. Hurley}
\author[2]{R.~Iaria}
\author[17]{D.~Kataria}
\author[18]{C.~Labanti}
\author[19]{G.~Lodato}
\author[20]{B.~Negri}
\author[21]{A.~Papitto}
\author[22]{T.~Piran}
\author[1]{A.~Riggio}
\author[23]{C.~Rovelli}
\author[24]{A.~Santangelo}
\author[25]{F.~Vidotto}
\author[17]{S.~Zane}
\affil[1]{Dipartimento di Fisica, SP Monserrato-Sestu, Universit\`a degli Studi di Cagliari, km 0.7, 09042 Monserrato, Italy} 
\affil[2]{Universit\`a degli Studi di Palermo, Dipartimento di Fisica e Chimica, via Archirafi 36, 90123 Palermo, Italy}
\affil[3]{INAF - Osservatorio di Astrofisica e Scienza dello Spazio di Bologna, Via Piero Gobetti 93/3, I-40129 Bologna, Italy}
\affil[4]{Dipartimento di Fisica Ettore Pancini, Universit\`a di Napoli ``Federico II'', and INFN, Sezione di Napoli, Complesso Univ. Monte S. Angelo, I-80126 Napoli, Italy}
\affil[5]{Gran Sasso Science Institute, I-67100 L`Aquila, Italy}
\affil[6]{Dipartimento di Fisica, Universit\`a di Napoli ``Federico II'', Complesso Universitario di Monte Sant`Angelo, Via Cinthia, 21, I-80126 Napoli, Italy}
\affil[7]{Dipartimento di Fisica ``G. Occhialini'', Universit\`a degli Studi Milano - Bicocca, Piazza della Scienza 3, Milano, Italy}
\affil[8]{INAF-IAPS,Via del Fosso del Cavaliere 100, I-00133 Rome, Italy}
\affil[9]{Physics Department, Universit\`a di Roma La Sapienza, Ple. Aldo Moro 2, 00185, Rome, Italy}
\affil[10]{Capodimonte Observatory, INAF-Naples , Salita Moiariello 16, 80131-Naples, Italy}
\affil[11]{Department of Astrophysics/IMAPP, Radboud University, P.O. Box 9010, 6500 GL Nijmegen, The Netherlands}
\affil[12]{INAF-IAPS,Via del Fosso del Cavaliere 100, I-00133 Rome, Italy}
\affil[13]{INAF-OATs Via G.B. Tiepolo, 11, I-34143 Trieste, Italy}
\affil[14]{Department of Physics and Earth Science, University of Ferrara, via Saragat 1, I-44122, Ferrara, Italy}
\affil[15]{INAF - Osservatorio Astronomico di Brera, via E. Bianchi 46, I-23807, Merate, Italy}
\affil[16]{University of California, Berkeley, Space Sciences Laboratory, 7 Gauss Way, Berkeley, CA 94720-7450, USA}
\affil[17]{Mullard Space Science Laboratory, University College London, Holmbury St. Mary, Surrey, RH5 6NT, UK}
\affil[18]{INAF-OAS Bologna, Via Gobetti 101, I-40129 Bologna, Italy}
\affil[19]{Dipartimento di Fisica, Universit\`a degli Studi di Milano, via Celoria 16, 20133 Milano, Italy}
\affil[20]{Italian Space Agency - ASI, Via del Politecnico, 00133 Roma, Italy}
\affil[21]{INAF-Osservatorio Astronomico di Roma, via Frascati 33, I-00040, Monteporzio Catone, Roma, Italy}
\affil[22]{Racah Institute of Physics, The Hebrew University of Jerusalem, Jerusalem 91904, Israel}
\affil[23]{CPT, Aix-Marseille Universit\`e, Universit\`e de Toulon, CNRS, Marseille, France}
\affil[24]{University of T\"{u}bingen-IAAT, Sand 1, D-72076 T\"{u}bingen, Germany}
\affil[25]{Department of Applied Mathematics, University of Western Ontario, London, ON N6A 5B7, Canada}
\affil[26]{INAF, Viale del Parco Mellini 84, I-00136 Roma, Italy}
\affil[27]{INFN, Sezione di Cagliari, Cittadella Universitaria, 09042 Monserrato, CA, Italy}

\authorrunning{Burderi et al.} 

\institute{ \noindent L. Burderi \at
           Dipartimento di Fisica, SP Monserrato-Sestu, Universit\`a degli Studi di Cagliari, km 0.7, 09042 Monserrato, Italy \\
              Tel.: +39-070-6754854\\
              \email{burderi@dsf.unica.it}           
           \and
           A. Sanna \at
           \email{andrea.sanna@dsf.unica.it}
}

\date{Received: date / Accepted: date}

\maketitle

\begin{abstract}


\gq\ (Gamma Ray Astronomy International Laboratory for QUantum Exploration of Space--Time) is a mission concept based on a constellation (hundreds/thousands) of nano/micro/small--satellites in low (or near) Earth orbits. Each satellite hosts a non--collimated array of scintillator crystals coupled with Silicon Drift Detectors with broad energy band coverage (keV--MeV range) and excellent temporal resolution ($\leq 100$ nanoseconds) each with effective area $\sim 100 \; {\rm cm}^2$. This simple and robust design allows for mass--production of the satellites of the fleet. This revolutionary approach implies a huge reduction of costs, flexibility in the segmented launching strategy, and an incremental long--term plan to increase the number of detectors and their performance: a living observatory for next--generation, space--based astronomical facilities. \gq\ is conceived as an all--sky monitor for fast localisation of high signal--to--noise ratio transients in the X/gamma--ray band, e.g.  the elusive electromagnetic counterparts of gravitational wave events. Robust temporal triangulation techniques will allow unprecedented localisation capabilities, in the keV--MeV band, of a few arcseconds or below, depending on the temporal structure of the transient event. The ambitious ultimate goal of this mission is to perform the first experiment, in quantum gravity, to directly probe space--time structure down to the minuscule Planck scale, by constraining or measuring a first order dispersion relation for light {\it in vacuo}. This is obtained by detecting delays between photons of different energies in the prompt emission of Gamma--ray Bursts.

\keywords{ constellation of satellites \and quantum gravity \and Gamma--ray Bursts   
\and $\gamma$--ray sources \and all--sky monitor}
\end{abstract}

\section{Introduction: Was Zeno right? -- A brief summary of Quantum Gravity and the in--depth structure of Space and Time}
\label{intro}
According to Plato, the great Greek philosopher, around 450 BC Zeno and Parmenides, disciple and founder of the
Eleatic School, visited Athens \citep[][{\it Parmenides}]{Plato} and encountered Socrates, who was in his twenties.
On that occasion Zeno discussed his world famous paradoxes, "four arguments all immeasurably subtle and profound",
as claimed by Bertrand Russell~\citep{Russell}.

In essence, Zeno's line of reasoning used, for the first time, a powerful logical method, the so-called
{\it reductio ad absurdum}, to demonstrate the logical impossibility of the endless division
of space and time in the physical world.

Indeed,  in his most famous paradox, known as {\it Achilles and the tortoise}, Zeno states that, if one admits as true the endless
divisibility of space, in a race the quickest runner can never overtake the slowest, which is patently absurd, thus
demonstrating that the original assumption of infinite divisibility of space is false.

The argument is as follows: suppose that the tortoise starts ahead of Achilles; in order to overtake the tortoise,
in the first place Achilles has to reach it. In the time that Achilles takes
to reach the original position of the tortoise, the tortoise has moved forward by some space, and therefore, after
that time, we are left with the tortoise ahead of Achilles (although by a shorter distance).
In the second step the situation is the same, and so on, demonstrating that Achilles cannot even reach the tortoise.

Despite the sophistication of logical reasoning, today we know that the error in the reasoning of Zeno
was the implicit assumption that an infinite number of tasks (the infinite steps that Achilles has to cover to reach the
tortoise) cannot be accomplished in a finite time interval, which is not true if the infinite number of time intervals
spent to accomplish all the tasks constitute a sequence whose sum is a convergent mathematical series.

However, the line of reasoning reported above exerts a certain fascination on our brains, which reluctantly accept the
fact that, in a finite segment, an infinite number of separate points may exist.

The mighty intellectual edifice of Mechanics developed by Newton has its foundations on
the convergence of mathematical series which serves to define the concept of the derivative
(fluxions, to use the name originally proposed by Newton for them), that are ubiquitous in physics.
Classical Physics has this idea rooted in the postulate (often implicitly accepted) that the physical quantities
can be conveniently represented and gauged by real numbers.

At the beginning of the last century, the development of Quantum Mechanics
revolutionised this secular perspective. Under the astonished eyes of experimental physicists,
Nature acted incomprehensibly when investigated at microscopic scales.
It was the genius of Einstein who fully intuited the immense intellectual leap that our minds were obliged
to accomplish to understand the physical world.
In a seminal paper of 1905~\citep{EinsteinI} the yet unknown clerk at the Patent Office in Bern
shattered forever the world of Physics by definitely proving, with an elegant explanation of Brownian motion,
that matter is not a continuous substance but is rather constituted by lumps of mass that were dubbed
Atoms by the English physicist Dalton in 1803. The idea that matter is built up by adding together minuscule
indivisible particles is very old, sprouting again from a surprising insight of Greek philosophers.
The word itself, {\it Atom}, which literally means indivisible, was coined by the ancient Greek philosophers
Leucippus and Democritus, master and disciple, around 450 BC, in the same period in which Zeno was
questioning the endless divisibility of space and time!

In 1905 Einstein completed the revolution in the physics of the infinitely small by publishing
another milestone of human thought~\citep{EinsteinII}  in which he argued that light is composed of
minuscule lumps of energy that were dubbed photons by the American physicist Troland in 1916.

The idea that the fundamental "bricks" of matter were indivisible particles with universal properties
characterising them like mass and electrical charge progressively settled into the physics world thanks to the
spectacular discoveries of distinguished experimental physicists. In a quick overview of this hall of fame
we have to mention (without claiming to perform a comprehensive review) Thomson, who discovered
the electron in 1896, Rutherford, who discovered in 1909 that the positive charge of the Atom was
concentrated in a small central nucleus, and discovered the proton in 1919, Chadwick, who discovered
the neutron in 1932, Reines, who discovered the neutrino in 1956, following Pauli who in 1930 postulated
its existence, Gell-Mann and Zweig, who proposed the existence of the quark in 1964, Glashow, Salam
and Weinberg, who proposed the existence of the W and Z gauge bosons in 1961, discovered by Rubbia and
van der Meer in 1983, and finally Higgs, Brout, Englert, Guralnik, Hagen, and Kibble who postulated the existence of the Higgs
boson in 1964, discovered at the CERN laboratories in 2011 by teams led by Gianotti and Tonelli.

Summarising, by the beginning of the third millennium physicists have developed and experimentally verified
a quite coherent and theoretically robust picture of the world at small scale that they dubbed with the rather
unprepossessing expression the Standard Model of Particle Physics, where the central role of the indivisible
fundamental bricks that build up the world is alluded into the word "Particle". After 2,500 years, the
formidable intuition of Greek philosophers has been confirmed: Democritus was right!

But what about Zeno? The mighty and flawless edifice of Calculus, developed
by giants of human thought like Archimedes, Newton and Leibniz, and the
elegant and audacious construction of Cantor, who demonstrated that even the endless divisibility
of fractional numbers was not powerful enough to describe
the immense density of real numbers -- and the name "real", used by mathematicians to
describe this type of number, alludes to the idea that they are essential to adequately gauge
the objects of the physical world -- seemed to have finally relegated the sophisticated logical
arguments of the philosopher from Elea to the endless graveyard of misconceptions.

However, the inverse square law,  the universal law discovered by Newton for gravitation,
that was successfully extended by Coulomb to the realm of electricity, and effectively
generalised by Yukawa in 1930 for a massive scalar field, contained the seed that would
resurrect the old proposal of Zeno in the vivacious crowd of modern scientific thought.

The crucial point is that the combination of the indivisible discreteness of some fundamental
properties, like mass or charge -- that allowed the development of the very concept of {\it elementary
particle}, cornerstone of the Quantum Field Theory, the mathematical formulation behind the
Standard Model -- is at odds with the generalised Yukawa potential widely used at least in the lowest
order formulation of the interaction of a pair of fermions in Quantum Field Theory. The crucial role of the
Yukawa potential in the development of Quantum Field Theory is evident when
using Feynman Diagrams (firstly presented by Feynman at the Pocono Conference in 1948) to represent the
interaction of a pair of fermions. In simple words, the Yukawa potential is divergent with $r \rightarrow 0$ and therefore
in contrast with the existence of point--like particles.

In our opinion the essence of the conflict between the "granular" world of Quantum Particles (excited states of the fields) and the continuum manifold that is used to represent the Minkowski Space--Time over which the fields are represented
has to be ascribed to the difficulty to insert, in the same logical scheme, the indivisible nature of elementary particles and the infinite divisibility of Space--Time over which Quantum Fields are defined.

To fully grasp this important aspect we must quickly summarise the stages through which the Fields, and the Space--Time on which they are defined, have become ``actors'' on the stage of physics playing an active supporting role, if not a dominant one, with respect to that of the Particles just discussed.

Together with Quantum Mechanics, General Relativity radically changed our understanding of Space and Time. According to the great philosopher Immanuel Kant, both these quantities are necessary {\it a priori} representations that underlie all other intuitions.
Indeed, in his Critique of Pure Reason, Kant says: {\it ``Now what are space and time? Are they actual entities? Are they only determinations or also relations of things, but still such as would belong to them even if they were not intuited? Or are they such that they belong only to the form of intuition, and therefore to the subjective constitution of our mind, without which these predicates could not be ascribed to any things at all?''}
These fundamental issues, raised by the German philosopher, outline the sense of the immense epistemological revolution bravely fought by the audacious physicists of the nineteenth and twentieth centuries.
Indeed, the seminal work of Maxwell and Einstein, just to mention the most prominent actors, has revealed that (electromagnetic) fields, space, and time, are not {\it a priori} categories of human thought, but physical objects, susceptible to experimental investigation.
Their physical properties would have turned out, in the years to come, to be very different from those that our intuition could suggest to us.
The initial albeit crucial point of this investigation can be identified in Maxwell's proposal of adding the ``displacement current'' term to one of the electromagnetic laws, already proposed by Coulomb, Faraday, and Amp\`ere.
The addition of this term determines a complete feedback of the electric and magnetic fields, in the absence of charges or currents, and, therefore, determines a physical reality for electromagnetic fields, that is independent of the presence of the charges, and currents that generated them.
Fields are no longer convenient mathematical tools to compute the forces acting on particles, but constitute physical objects endowed with their own independent existence!
From the wave equation implied by these new laws, Maxwell obtained the constant that expresses the speed of propagation of these fields with respect to the vacuum.
The genius of Einstein understood that the combination of the constancy of the speed of light with the principle of relativity, proposed in 1632 by Galilei in his Dialogue on the two greatest systems of the world, was to unhinge our Newtonian conception of absolute Space and Time, independent of each other. This led him to the extraordinary conception of a deformable Space--Time, subject to the constraint of Lorentzian invariance.
However, the price to pay for this epistemological revolution, was the acknowledgement that, operationally -- in the Bridgmanian sense of the term~\citep{Bridgman27} -- it is impossible to synchronise the clocks, and/or to define the distances, in an instantaneous way or, in any case, faster than imposed by the speed of light in vacuum.
This led Einstein to the intuition that also Gravity (the only other field known at the time) should propagate through a wave equation, at the same speed determined by Maxwell's equations.
Indeed, in their weak field limit, the field equations of General Relativity resemble Maxwell's equations, in the presence of the so--called Gravito-magnetic Field, a field generated by matter currents, in perfect analogy with the Magnetic Field generated by charge currents.
Again, through the complete feedback determined by the equations relating temporal and spatial variations of Gravitational and Gravito-magnetic Fields, a wave equation was capable of describing the propagation of Gravitational Fields through the vacuum, at the very same speed as the Electromagnetic Fields!
The overall coherence of this epistemological revolution, imposed by Special Relativity, was guaranteed by acknowledging that Space--Time was a physical entity, subject to oscillations in its texture, and not a couple of philosophical {\it a priori} categories, as discussed by Kant.

In summary, in modern physics, space and time have progressively changed their role. From mere passive containers of events (in line with the Kantian idea of mental categories) to physical quantities that, combined in the unique hyperbolic geometry implied by the constancy of the speed of electromagnetic waves, are able to deform under the gravitational action of the fields and of the particles. With due attention, the Space--Time of General Relativity can be considered, for all intents and purposes, a field with its associated quantum particles (excited states of the fields): the gravitons. In this unifying picture, macroscopic coherent states of a huge number of gravitons are the gravitational waves, recently detected by the LIGO and Virgo observatories.

The tension between the granularity of quantum particles and the continuity of fields (defined by real variables) has been alleviated by renormalisation techniques fully applicable in Gauge Theories of Quantum Field, as shown by Gerard 't Hooft for all fundamental forces except gravity.
Renormalisation techniques have proven to be extremely effective in solving the problem of the infinities that arise when, in Quantum Field Theory, we try to combine point--like particles with fields diverging for $r \rightarrow 0$.
This approach is based on the existence of ``charges'' of opposite sign capable of producing, in the calculations of the associated physical quantities, terms of opposite sign which, although diverging, cancel each other out, when treated with sufficient care.

Despite their success, renormalisation techniques seem to be inadequate when gravity comes into play.
Because of the mass-energy equivalence predicted by Special Relativity, the natural generalisation of the source ``charge'' of the gravitational field is the entire energy density and not only that associated with the rest mass of the particles.
This implies that any type of field attempting to prevent gravitational collapse acts, through the energy density (usually positive) associated with it, as a further source of gravitational field, preventing, in fact, an effective renormalisation. This last feedback is difficult to eliminate within the framework just described and makes clear, in our opinion, the conceptual stalemate that prevents, at the present time, the unification of the two most revolutionary physical theories of the twentieth century: General Relativity and Quantum Mechanics.
Indeed, a novel ingredient, peculiar to General Relativity, prevents the propagation, in the surrounding universe, of the oddities associated with a divergent field, by enshrouding the singularity with an {\it Event Horizon}, a surface on which time is frozen by the intensity of the gravitational grip. However, the formation of these Event Horizons around gravitational singularities is not guaranteed by the mathematical structure of the theory, in which singularities not surrounded by Event Horizons are dubbed {\it Naked Singularities}. In order to guarantee self--consistency of the whole picture, in 1969 Roger Penrose conceived the so-called {\it Cosmic Censorship Hypothesis}, that no naked singularities exist in the universe \citep{Penrose69}. Beside being an {\it ad hoc} conjecture, not stated in a completely formal way,
a lively scientific debate is currently underway regarding the validity of the proposed conjecture, e.g.  the somewhat related Thorne--Hawking--Preskill bet
\citep[{\it ``Black hole information bet''}, see e.g.][last chapter]{Susskind}.
In this perspective, Extended Theories of Gravity represent an approach to overcome the lack of a final theory of Quantum Gravity \citep[see e.g.][]{CapozzielloDeLaurentis2011}.

To overcome this formidable impasse, theoretical physics is today exploring more radical approaches that require a new conceptual revolution, {\it a paradigm shift}, to use Kuhn's words.

Here we just mention two opposite approaches that tackle the problem of the irresolvable dichotomy of particles and fields from somewhat opposite perspectives. String Theories \citep[see e.g.][for reviews and later criticism of this approach]{smolin2007} that eliminate the point--like nature of the particles by assigning to each of them a (mono)--dimensional extension: the string. Loop Quantum Gravity \citep[see e.g.][for reviews]{Rovelli98} which questions the smoothness of Space--Time, quantising it into discrete energy levels like those observed in classical quantum--mechanical systems to form a complex {\it pregeometric structure} (to use the words of Wheeler) dubbed Spin--Network.

Both proposed theories (although with different and somewhat opposite theoretical approaches) imply the existence of a minimal length for physical space (and time). The emergence of {\it Atoms of Space and Time} -- to use an efficacious and vivid expression, coined by Smolin in 2006 -- is a necessary consequence of the ultimate quantisation of Space--Time.

However the spatial (and temporal) length--scales associated with this quantisation, are minuscule, in terms of standard units, as already suggested in a pioneering and visionary work of Planck in 1899~\citep{Planck99}: $\ell_{\rm P} \sim \sqrt{\hbar G/ c^3} \sim 10^{-33}$ cm and $t_{\rm P} \sim \sqrt{\hbar G/ c^5} \sim 10^{-43}$ s for the Planck length and time, respectively.
For comparison, the shortest distance (Compton wavelength) directly measured to date at the Large Hadron Collider at CERN is $\sim 10^{-20}$ cm (for colliding energies of few $10^{12}$ eV). The shortest time intervals ever measured are just above atto-seconds $\sim 10^{-18}$ s \citep[see e.g.][]{Hentschel01}.
Experimentally, at the present moment, we are more than ten orders of magnitude above the theoretical limit we would like to probe to effectively constrain our theoretical speculations!

For a quick (and not exhaustive) overview of the variety of theoretical approaches exploring the possibility of the existence of fundamental limits in the ability to measure (and therefore to define, in the Bridgmanian sense) intervals of arbitrarily small space and time, we use, almost textually, what is reported in a recent work by some of us~\citep{Burderi16} and references therein.

Several thought experiments have been proposed to explore fundamental
limits in the measurement process of time and space intervals
\citep[see e.g.][for an updated and
complete review]{Hossenfelder12}. In particular Mead~\citep{Mead64} ``postulate the existence
of a fundamental length'' (to use his own words) and discussed the possibility
that this length is the Planck length, $\ell_{\rm min} \sim \sqrt{G \hbar/c^3} = \ell_{\rm P}$, which resulted in
limitations in the measurement of arbitrarily short time intervals giving rise to relations similar to the Space--Time Uncertainty relation proposed by~\citet{Burderi16}.
Moreover, in a subsequent paper  \citep{Mead64}, Mead discussed an -in principle-
observable spectral broadening, a consequence of the postulate of the existence
of a fundamental length of the order of the Planck Length.
More recently, in the framework of String Theory, \citet{Yoneya87,Yoneya89} proposed a space-time
uncertainty relation which has the same structure as the uncertainty relation
discussed in the aforementioned paper of \citet{Burderi16} \citep[see e.g.][for a discussion of
the possible role of a space-time uncertainty relation in String Theory]
{Yoneya97}.
The relation proposed in String Theory
constrains the product of the uncertainties in the time interval $c \Delta T$
and the spatial length $\Delta X_l$ to be larger than the square of the
string length $\ell_S$, which is a parameter of the String Theory.
However, to use the same words as Yoneya \citep{Yoneya97}, this relation is
{\it ``speculative and hence rather vague yet''}.
Indeed, in the context of Field Theories, uncertainty relations
between space and time coordinates similar to those proposed here
have been discussed as an {\it ansatz} for the limitation arising in combining
Heisenberg's uncertainty principle with Einstein's theory of
gravity \citep{Doplicher95}.
\citet{Garay95} postulated and discussed, in the context of Quantum Gravity,
the existence of a minimum length of the order of the Planck Length, but followed the idea
that this limitation may have a similar meaning to the speed limit defined by the speed
of light in Special Relativity, in line with what was already pointed out previously
\citep[see e.g.][and references therein]{Borzes88}.
In the framework of the so-called Quantum Loop Gravity \citep[see e.g.][for a review]{Rovelli88a,Rovelli88b,Rovelli98} a minimal length appears characteristically in the form
of a minimal surface area \citep{Rovelli95,Astekhar97}: indeed the area
operator is quantised in units of $\ell_{\rm P}^2$ \citep{Rovelli93}.
It has been sometimes argued that this minimal length might conflict with Lorentz invariance,
because a boosted observer could see the minimal length further Lorentz contracted.


Indeed, some of the proposed theories allow for this Lorentz Invariance Violation (LIV, hereinafter) at some small scales \citep[see e.g.][for reviews]{Mattingly05,Amelino09,Liberati13}. Essentially in these scenarios
the presence of a granular structure of space in which electromagnetic waves ({\it i.e.} photons, from the quantum point of view) propagate, determines the emergence of a dispersion law for light in vacuum, in close analogy with what happens for the propagation of photons in a crystal lattice.

We should stress that not all ways of introducing spacetime granularity will produce these dispersive effects. In particular,
in Loop Quantum Gravity the granularity is mainly reflected in a minimum value for areas which however, is not
a fixed property of geometry, but rather corresponds to a minimal (nonzero) eigenvalue of a
quantum observable that has the same minimal area $\ell_{\rm Planck}^2$ for all the
boosted observers (what changes continuously in the boost transformation is the probability distribution
of seeing one or the other of the discrete eigenvalues of the area \citep[see e.g.][]{Rovelli03}).
However, in Loop Quantum Gravity there are results amenable for testing with gamma-ray telescopes,
the most studied possibility being an anomalous dependence of frequency on distance, producing a flattening of
the cosmological redshift \citep{Barrau14}.

The energy scale at which dispersion effects become manifest can be easily computed e.g.  equating the photon energy, $E = h \nu$, to $\nu \sim 1/t_{\rm P}$ which provides the Planck Energy $E_{\rm P} \sim \sqrt{\hbar c^5/G} \sim 10^{28}$ eV, a huge energy for the particle's world, corresponding to the mass of a paramecium ($\sim 0.02$ mg).
Again, frustratingly, this energy scale is well beyond any possibility of direct investigation with any kind of colliders in the near and distant future.
It is worth noting that, in the simplest models, at lowest order, the dispersion law for the photon speed $v_{\rm phot}$ is dominated by the linear term: $\delta v_{\rm phot}/c \propto h\nu/\sqrt{\hbar c^5/G}$, with constant of proportionality
$\xi \sim 1$.

In our opinion, this unprecedented situation, in which the scale of the expected experimental phenomena is very far from the current possibilities of experimental verification,
is hampering any significant progress
in our understanding of the ultimate structure of the world.
Physics is, after all, an experimental discipline in which continuous comparison with experimental data is essential, even to draw unexpected clues from which to develop new theories. This was the case for the development of Relativity and Quantum Mechanics in which bold physicists and epistemologists had to develop new logical models to account for  unexpected experimental results that were unimaginable for the classical conception of nature developed by Greek philosophers. Indeed, the fatal blow to the classical conception of physics developed up to Newton and Maxwell, was given by the experimental impossibility to determine the speed of Earth with respect to the {\it Cosmic Aether} (the medium in which electromagnetic waves propagate) as firmly established by the null result of the Michelson and Morley experiment \citep{Michelson87}.

Indeed, in the context of Quantum Gravity, we are witnessing a flourishing of countless elegant mathematically daring theories, which testify the lively interest of brilliant minds towards problems of undoubted physical and epistemological relevance that sadly, at the moment, lack the invigorating and vitalising confrontation with constraining experimental data.

For comparison, the recent discoveries of the existence of the Higgs Boson, which confirmed and strengthened it, the Standard Model of Particle Physics, the detection of Gravitational Waves, which confirmed what was predicted a century ago by General Relativity and the recent spectacular image, obtained interferometrically, of the event horizon around a supermassive black hole, which confirmed the formation of trapped surfaces in the Space--Time fabric, have revitalised these very interesting fields of research by opening the doors to new disciplines such as Multi--Messenger Astronomy \citep{Abbot_mu_2017}.

However, we believe that a {\it giant leap} is now possible also in the difficult experimental task of investigating the texture of Space on the minuscule scales provided by Quantum Gravity.
In the following we will show how
the technological progress in Space Sciences and the enormous reduction in the costs necessary to put detectors into space, can allow us to conceive an ambitious experiment to verify, for the first time, directly, some of the most important consequences of the existence of a discrete structure for the texture of the space.
To put it suggestively, twenty-five centuries after the meeting of the Eleatic philosophers with Socrates in Athens, we are able to investigate the problem raised by Zeno in a quantitative way.

In particular, in line with the suggestions outlined in some pioneering works in the field of experimental investigation of Quantum Gravity \citep{Camelia98,Capozziello15}, we propose a ambitious albeit robust experiment to directly search for tiny delays in the arrival times of photons of different energies determined by the dispersion law for photons discussed above. Given the hugeness of the Planck Energy, we expect, as will be shown in \S \, \ref{delays}, delays $\sim$ few $\mu$s for Gamma--ray Burst
(GRB) photons that travelled for more than ten billion years!

These last numbers show, in themselves, the difficulty and ambitiousness of the proposed experiment. We would like to emphasize here, however, that even a null result, that is a solid proof of the non--existence of a linear effect in the law of photonic dispersion for energies normalised to the Planck scale, would constitute a result of capital importance for the progress of fundamental physics. After all, the aforementioned Michelson and Morley experiment, decisive for the acceptance, in an understandably conservative scientific community, of the revolutionary ideas on space and time implied by the Theory of Relativity, provided a null result with respect to the possibility of identifying motion with respect to
the {\it Cosmic Aether}!


A promising method for constraining a first order dispersion relation for photons {\it in vacuo} is the study of discrepancies in the arrival times of high-energy photons of Gamma-Ray Bursts (sudden and unpredictable
bursts of hard-X-$\gamma$ rays, with huge fluxes up to $10^{-2}$
ergs/cm$^{2}$/s, emitted at cosmological distances, GRB hereafter) in different energy bands. Despite the relevant number of papers, published in recent years \citep[see e.g.][for a comprehensive analysis of Fermi--LAT gamma-ray burst data]{Ellis19}, we believe that the first order dispersion relation has not yet been investigated with the due accuracy because, at present, we lack an experiment with all the desired characteristics to effectively constrain this relation, beyond any possible loophole.
In particular, our major concerns are possible intrinsic delays (characterising the emission process) superimposed over the tiny quantum delays. This is particularly evident in the caveat discussed in \citet{Abdo09b} on GRB090510 and, more recently, in the paper by \citet{Wei17} and \citet{Ellis19} who set a robust constraint on LIV using Fermi--LAT GRB data of a few $10^{17}$ GeV. Further indications of no LIV violations come from the HESS collaboration, in particular from spectral analysis of the blazar Mrk 501 \citep{Lorentz17}, although also in this case a spectral shape and hypothesis on the emission process are assumed. Moreover, all these analyses assume a dependence of the effects on redshift which was conjectured in the pioneering paper by \citet{Jacob08}; however as theorists acquire the
ability to test the Jacob-Piran conjecture in explicit models it is often found that other forms of redshift dependence apply \citep[see e.g.][]{Rosati15}.
In our opinion, given the importance of the question, a direct robust measurement cannot be based on the analysis of a single object and a robust statistical analysis of a rich sample of data is required, in which the natural direct timescale of the LIV-induced delays in the gamma-ray band (one microsecond) is thoroughly searched.
None of the experiments discussed above had the right combination of time resolution and collecting area to effectively scrutinise this regime.



\section{\gq\ and its scientific case in a nutshell}
The coalescence of compact objects, neutron stars (NS) and black holes
(BH), and the sudden collapse to form a supra--massive NS or a BH, hold the keys to
investigate both the physics of matter under extreme conditions, and
the ultimate structure of space--time. At least three main discoveries
in the past 20 years prompted such studies.

Prompt arcminute localisations of GRBs enabled by the instruments on board
{\it BeppoSAX}, allowed the discovery of their X--ray and optical afterglows
\citep{Costa97,VanParadijs97}, which led to the
identification of their host galaxies \citep{Metzger97}. This definitely
confirmed the extragalactic nature of GRBs and assessed their energy
budget, thus establishing that they are the most powerful accelerators
in the Universe. Even accounting for strong beaming, the energy
released can indeed attain $10^{52-53}$ erg, a large fraction of the
Sun's rest mass energy, in $\approx 0.1-100$ seconds, produced by the bulk
acceleration of plasmoids to $\Gamma\approx100-1000$ \citep{Bloom09,Abdo09b}.

Second, the large area telescope (LAT) on board the Fermi satellite
confirmed GRBs as GeV sources as previously reported by the {\it EGRET} instrument on board the Nasa Compton Gamma-Ray Observatory satellite, confirming their capability to
accelerate matter up to $\Gamma\approx100-1000$ and allowing us to
apply for the first time the program envisioned by Amelino-Camelia and
collaborators at the end of the 90's \citep{Camelia98} to investigate quantum
space-time using cosmic sources.

Third, the recent discoveries of the gravitational wave signals from several
BH--BH and one NS--NS mergers by Advanced LIGO and Virgo \citep{GW150914,GW170104,GW170814},
opened a brand new window to investigate the astrophysics of compact
objects, as well as fundamental physics. The gravitational signal
carries a huge amount of information on the progenitors and final
compact objects (masses, spins, luminosity, distance etc.). Moreover,
the current values for the number of mergers (rate in excess
of 12 Gpc$^{-3}$yr$^{-1}$), implies that the number of Gravitational Wave Events (GWEs hereafter)
associated with the merging of two compact objects is significant.

These scenarios and limits will be further constrained and improved in the
coming few years when the sensitivity of the interferometers will be
further improved, and the corresponding volume for BH--BH and NS--NS merging events
further enlarged. The activation of a third interferometer,
Advanced Virgo, on August 2017, has already greatly improved the
localisation capability of the Advanced LIGO/Virgo system, producing
error boxes with areas of a few hundreds of deg$^2$, 10-100 times smaller than those
provided by Advanced LIGO \citep{GW170814}. The localisation accuracy
will reduce to a few tens of deg$^2$ with the advent of KAGRA.

In August 2017 the first NS--NS merging event was discovered by LIGO/Virgo \citep{GW170817},
with an associated short GRB seen off-axis and detected first by the Fermi gamma-ray burst monitor (GBM),
{\it INTEGRAL}/SPI-ACS \citep{GW170817_GW}, and, only nine days after the prompt
emission, by Chandra
\citep{Troja17}. The GBM provided a position with uncertainty
$\sim12$ deg (statistical, 1$\sigma$, to which a systematic
uncertainty of several deg should be added).
The LIGO/Virgo error boxes led to the first identification of an optical transient associated with both a short GRB and a GWE, opening, {\it de facto}, the window of
multi--messenger astrophysics \citep{Abbot_mu_2017}.
This exciting new field of astrophysics research will allow us, in the immediate future, to obtain physical and cosmological information of paramount importance for our understanding of the GWE and GRB phenomena \citep[see e.g.][]{Phinney09}.

These considerations show that, in the near future, the prompt accurate localisation of the possible transient electromagnetic counterparts of GWEs is mandatory in order to fully exploit the power of scientific investigation of Multi-messenger Astronomy. Indeed, a high sensitivity to transient events in the X--ray/Gamma--ray window and their subsequent fast localisation with accuracies in the arcminute range or below, are mandatory in order to point narrow field instruments to scrutinise the GWE's electromagnetic counterparts in the whole electromagnetic band.

In addition, as discussed in Section \ref{intro}, GRB light--curves in different energy bands, in the X-ray/gamma-ray window, with temporal resolution $\le 1 \mu{\rm s}$, can be used to investigate a dispersion law for photons, predicted in some of the proposed theories of Quantum Gravity.

In summary, there are at least three broad areas that can and must be tackled in the next few years:

\begin{enumerate}
\item
the accurate (arcminute/arcsecond) and prompt (seconds/minutes) localisation of
bright transients;
\item
the study of the transients' hard X--ray/Gamma--ray temporal variability (down to the
microsecond domain and below, i.e three orders of magnitude better
than the best current measurements), as a proxy for the physical activity of the so-called {\it inner engine} that powers the most powerful explosions in our Universe;
\item
the use of fast high energy transients to investigate the structure of
space-time.
\end{enumerate}

We will discuss these three broad themes in the next Sections. We
devote the last Sections to a description of our proposed approach to
tackling the three main science themes listed above;
this consists of a distributed
instrument, a swarm of simple but fast hard X--ray/Gamma--ray detectors hosted by
small/micro-satellites in low Earth orbit, the \gq\ mission, specifically conceived
to provide precise measurements of the three main scientific themes mentioned above.

\section{Gamma-Ray Burst simulations and timing accuracy in Cross--Correlation analysis}
 
\subsection{Gamma--ray Burst fast variability}
 
GRBs are thought to be produced by the collapse of massive stars
and/or by the coalescence of two compact objects. Their main
observational characteristics are their huge luminosity and fast
variability, often as short as one millisecond, as shown by 
\citet{Walker00}, both in isolated flares and in lower amplitude
flickering. These characteristics soon led to the development of the
{\it fireball model}, i.e. a relativistic bulk flow where shocks
efficiently accelerate particles. The cooling of the
ultra-relativistic particles then produces the observed X-ray and
$\gamma$-ray emission. One possibility to shed light on their inner 
engines is through GRB fast variability.  Early numerical simulations 
\citep{Kobayashi97,Ramirez00,Spada00} suggested that the GRB light-curve 
reproduces the activity of the inner engine. More recently, hydrodynamical simulations of GRB jets showed that, in order to reproduce the 
observed light-curves, fast variability must be injected at the base of 
the jet by the inner engine, while slower variations may be due to the 
interactions of the jets with the surrounding matter \citep{Morsony10}.
 
The most systematic searches for the shortest timescales in GRBs so
far are those of \citet{Walker00}, \citet{MacLachlan13} and
\citet{Bhat12}. The first two works exploit rather sophisticated
statistical (wavelet) analyses, while the latter performs a parametric
burst deconvolution into pulses. \citet{Walker00} conclude that the 
majority of analysed {\it BATSE} GRBs show rise-times faster than 4 msec and 
30\% of the events have rise-time faster than 1 msec (observer frame).  
\citet{MacLachlan13} use Fermi/GBM data binned at 200$\mu$s (the accuracy of the GBM  time tagging is 2$\mu$s) and report somewhat longer minimum 
variability timescales than \citet{Walker00}, but conclude that variability 
of the order of a few msec is not uncommon (although they are limited by 
the wider temporal bin size adopted of 200$\mu$s and much worse statistics 
with respect to the {\it BATSE} sample).  Systematically longer time-scales are reported 
by \citet{Bhat12}, using data binned at 1 msec. This is not surprising, 
because direct pulse deconvolution requires the best statistics, which can hardly 
be obtained for the shortest pulses.

\subsection{Synthetic Gamma-Ray Bursts}
\label{sec:sim_des}

To estimate the accuracy obtainable from cross-correlation analysis, $E_{CC}$, 
defined as the standard deviation $\sigma$ of the distribution of delays obtained 
applying cross-correlation techniques to pairs of simulated GRB light curves,
we started by creating synthetic Long and Short GRBs with the following 
characteristics. The Long and Short GRBs considered have durations 
$\Delta t_{\rm Long} = 25 \; {\rm s}$ and $\Delta t_{\rm Short} = 
0.4 \; {\rm s}$, respectively.
To simulate the GRB's variability with a time-scale of $\sim 1\; {\rm ms}$ 
we assumed that each GRB results from the superposition of a great number 
of identical exponential shots of decay constant $\tau_{\rm shot} \sim 1 \, {\rm ms}$,
randomly occurring at an average arrival rate of $\lambda_{\rm shot} = 100 \; 
{\rm shot/s}$ for the entire duration of the GRB. 
The amplitude of each exponential shot is normalised to have a flux of 
$8.0 \; {\rm counts/s/cm^2}$ in the energy band $50 \div 300 \;{\rm keV}$, 
while the background photon flux in the same energy band has been fixed to 
$2.8 \; {\rm counts/s/cm^2}$ (consistent with typical backgrounds observed by Fermi GBM).
 
 \begin{figure}[h]
 \centering       
    \includegraphics[scale=0.28]{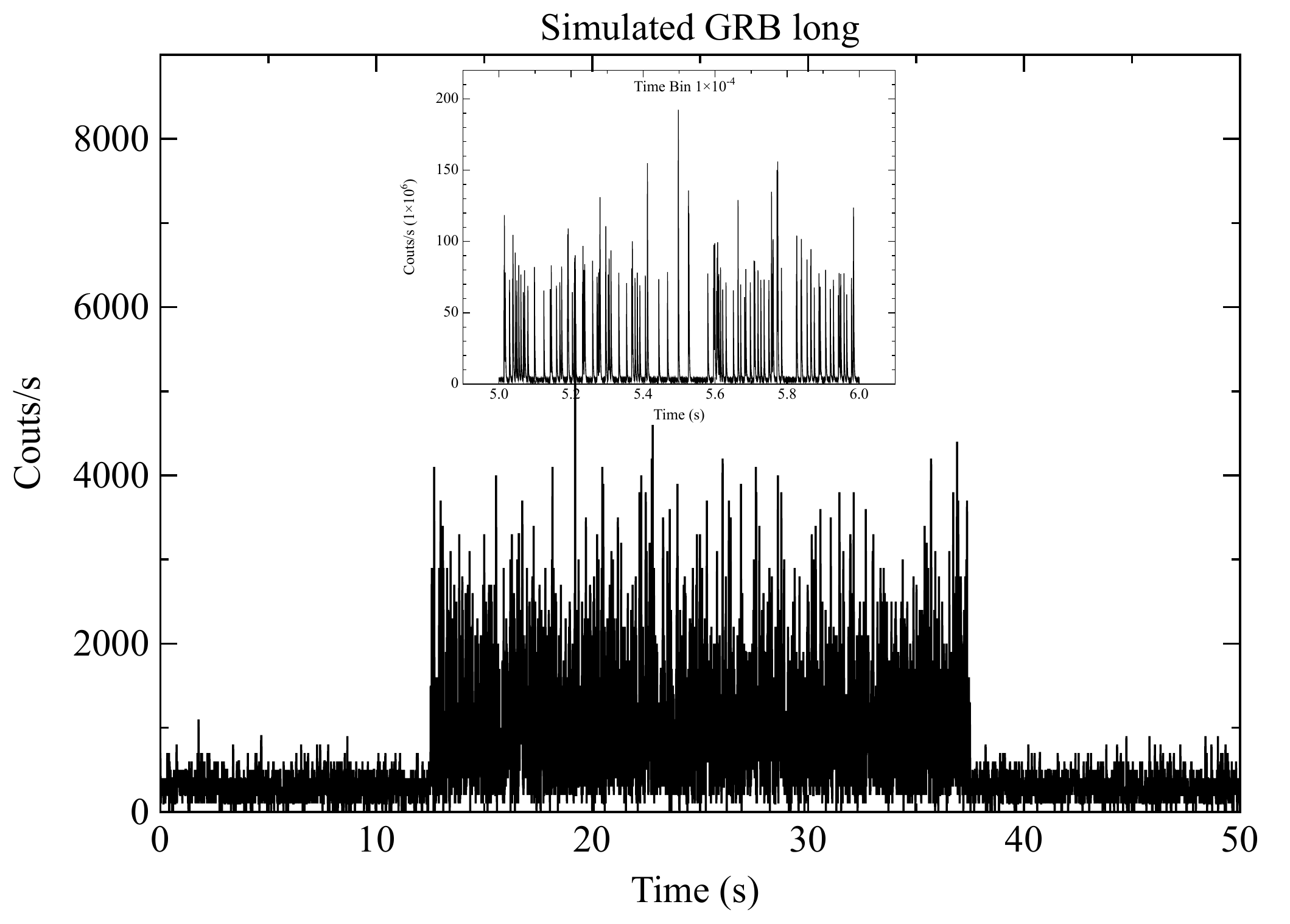}
    \hspace{2px}
    \includegraphics[scale=0.28]{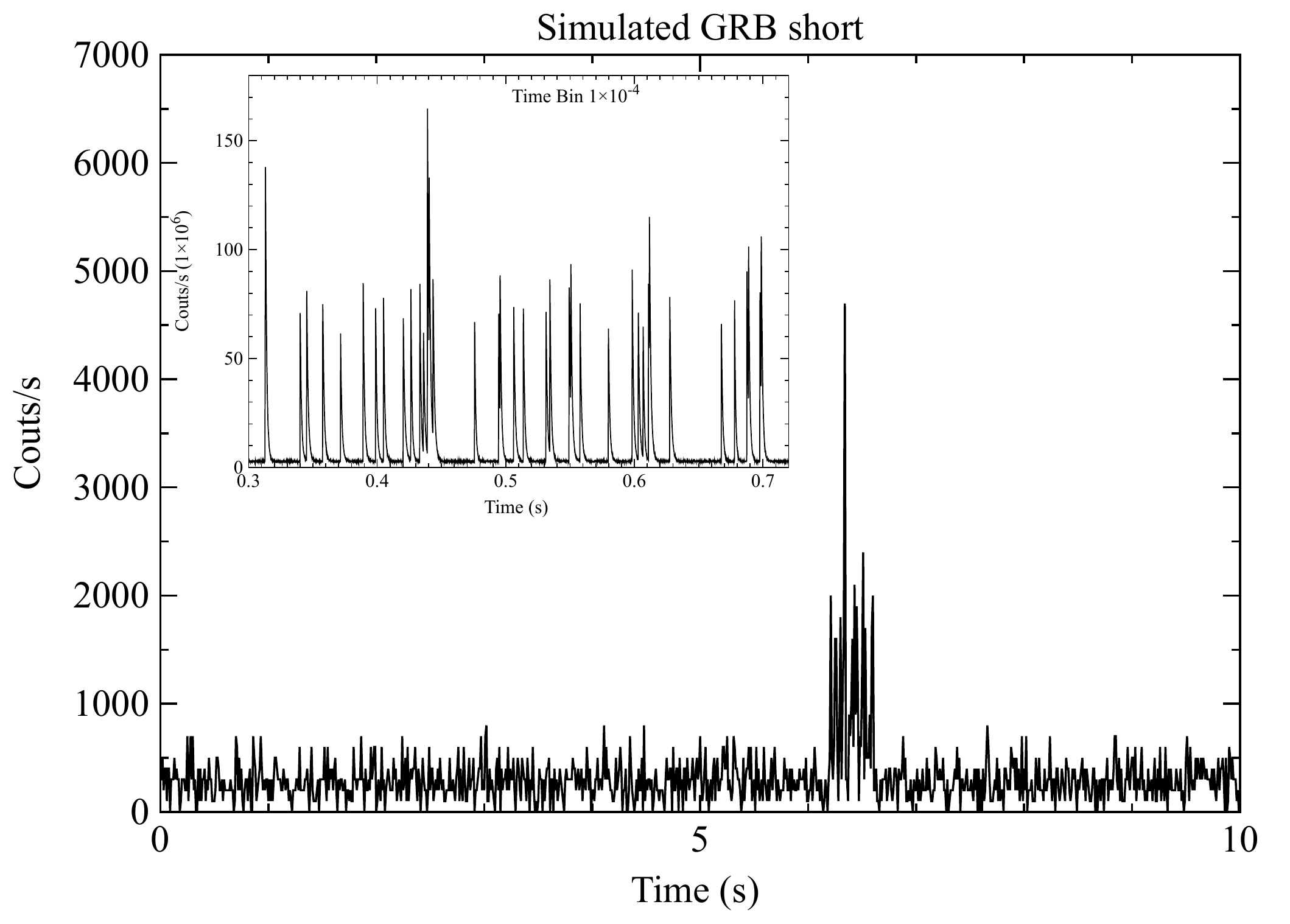}
    \caption{Light curves on timescales of $10^{-2}$ seconds for the synthetic long (top panel) and short (bottom panel) GRBs created following the procedure described in Sec.~\ref{sec:sim_des}. The insets show a zoom-in of the light curves created on shorter timescales ($10^{-4}$ seconds) after rescaling the effective area of the equivalent detector up to 100 square meters. }
    \label{fig:sim}
\end{figure}
Fig.~\ref{fig:sim} shows the synthetic light curves for the long (top panel) and 
short (bottom panel) GRBs, respectively, calculated accumulating photons on time 
scales of $10^{-2}$ seconds. The simulated GRB millisecond variability can be 
inspected in greater detail in the insets on Fig.~\ref{fig:sim}, in which a small 
fraction of the same light curves has been simulated increasing the equivalent 
effective area of the detector up to 100 square meters and accumulating photons on 
timescales two orders of magnitude shorter ($10^{-4}$ seconds).

\subsection{Fermi GBM Gamma--ray Bursts}
\label{sec:grbsel}
\begin{figure}[h]       
    \centering
    \includegraphics[scale=0.3]{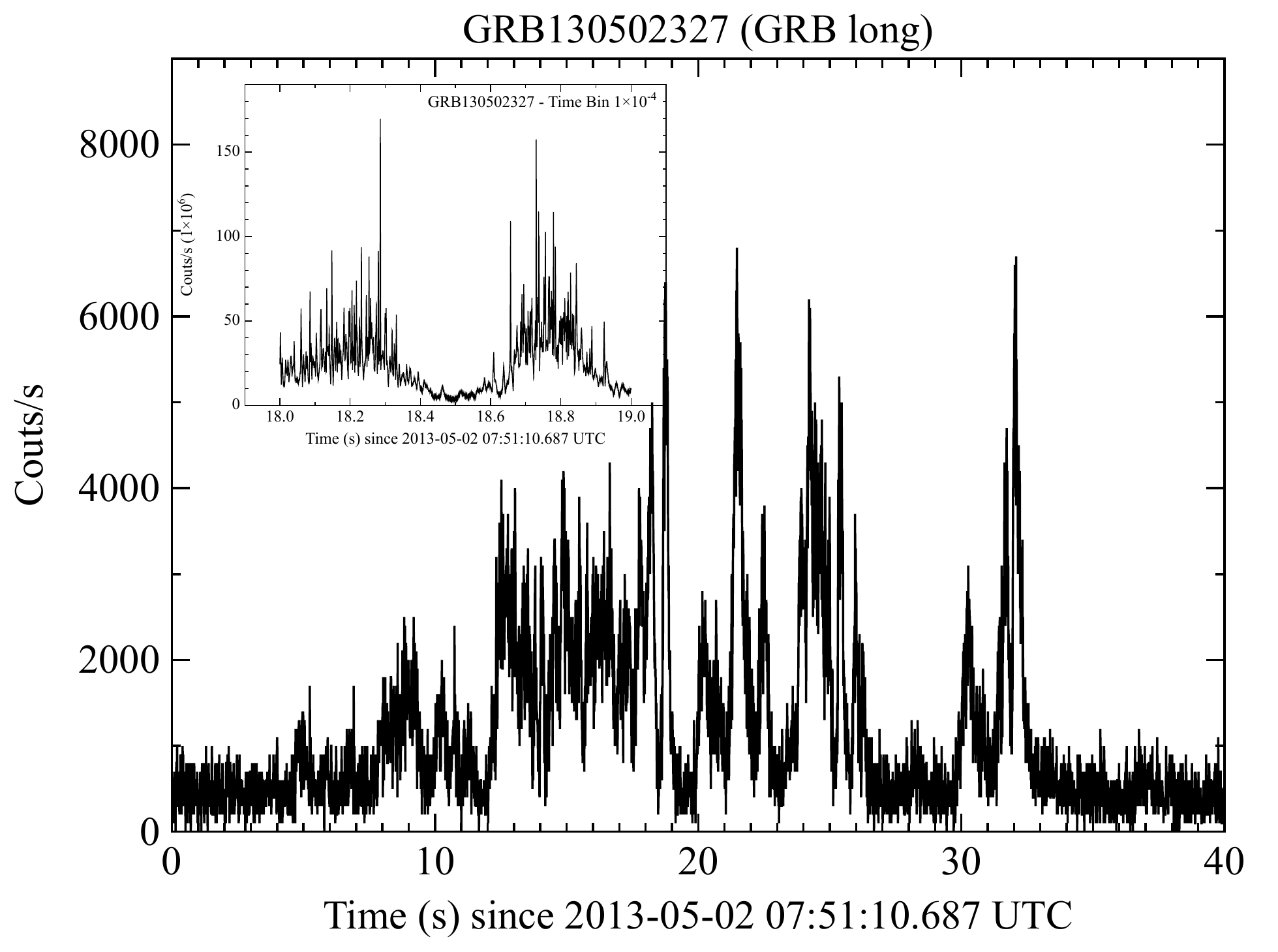}   
    \hspace{2px}
    \includegraphics[scale=0.3]{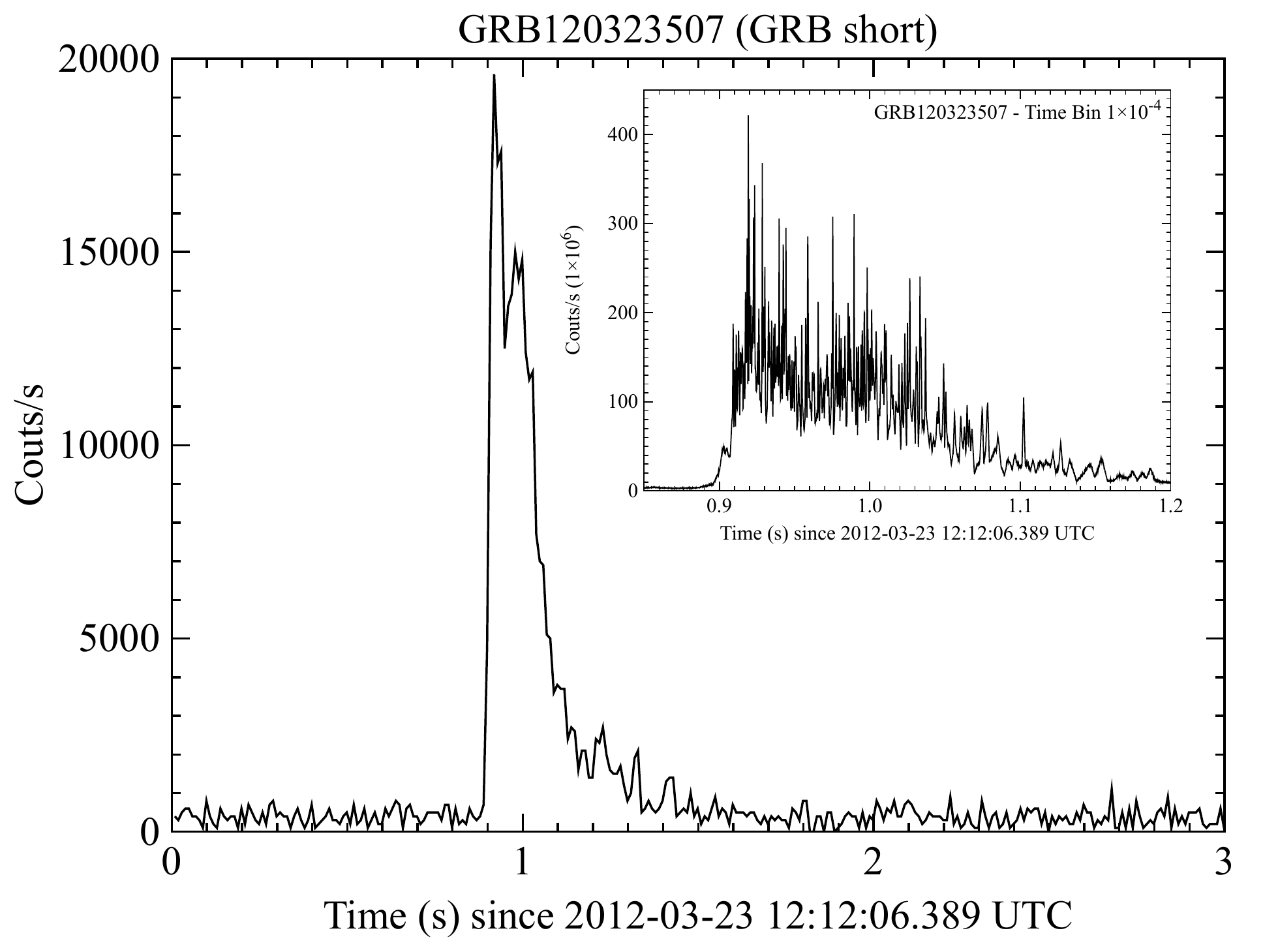}
    \caption{Light curves on timescales of $10^{-2}$ seconds for Long (top panel) 
and Short (bottom panel) GRBs detected by Fermi GBM (see Sec.~\ref{sec:grbsel} 
for more details). The insets show a zoom in of the simulated 
light curves created on shorter timescales ($10^{-4}$ seconds) after rescaling 
the effective area of the equivalent detector up to 100 square meters.}
    \label{fig:gbm}
\end{figure}

To further investigate the method we applied the same techniques to real data. 
In order to achieve the objectives extensively described above, we performed 
Monte-Carlo simulations based on real detections of GRBs obtained with GBM. We searched 
the available Fermi GBM archive seeking GRB's characterised by variability 
on time scales as short as a few milliseconds in order to enhance the sensitivity 
of time delay measurements between photons of different energies as well as the 
localisation of the GRBs prompt emission. 
For this work we selected the following events: a) a Short GRB (GRB120323507) 
observed on 2012 March 23, characterised by a $t_{90}$\footnote{This parameter 
represents the duration, in seconds, during which the 90\% of the burst fluence 
was accumulated.} duration of $\sim0.4$ seconds with a fluence of 
$\sim1\times 10^{-5}$ erg/cm$^2$; b) a Long GRB (GRB130502327) observed on 2013 
May 2, characterised by a $t_{90}$ duration of $\sim24$ seconds and a fluence of 
$\sim1\times 10^{-4}$ erg/cm$^2$. 
Fig.~\ref{fig:gbm} shows the light curves of the two selected events accumulated on 
$10^{-2}$ seconds timescales.

Simulations on short time scales ($\sim0.1$ ms) of a unique type of transient 
event such as a GRB, based on observed light curves, can be challenging when the 
effective area of the detector is so small that the statistics are fully dominated 
by Poissonian fluctuations that unavoidably characterise the (quantum) detection 
process. 
In particular, if the detected counts within the given time scale is $\le 1$, 
quantum fluctuations of the order of $100\%$ are expected. If, naively, the 
number of counts per bin is simply rescaled to account for an increase of 
effective area, these quantum fluctuations can introduce a false imprint of 
$100\%$ variability with respect to the original signal. No definite cure is 
available to mitigate this problem, that could be, however, alleviated by 
rebinning and/or smoothing techniques. Although smoothing techniques allow us to 
create light curves for any desired temporal resolution, correlations between 
subsequent bins is unavoidable. Cross-correlation techniques are strongly 
biased by this effect, hence we opted for a more conservative method involving 
standard rebinning in which the number of photons accumulated in each (variable) 
bin is fixed. After several trials and Monte-Carlo simulations we find that 
6 photons per bin allows us to preserve the signal variability while introducing undesired 
fluctuations not larger than $\sim 30\%$. 
Applying this rebinning technique to the GBM light curves (at the maximum time 
resolution of $2\mu$s) discussed above, we generated a variable bin size light 
curve. In order to produce a template for Monte-Carlo simulations, usable on any 
time scale, we linearly interpolated the previous light curve to create a 
functional expression (template) for the theoretical light curve. We note explicitly 
that linear interpolation between subsequent bins is the most conservative approach 
that does not introduce spurious variability on any time scale.
 
For a given temporal bin size, we amplified the GRB template previously described 
in order to take into account the overall effective area of the detector(s) and 
used this value as the expectation number of photons within the bin. 
Poissonian randomisation was then applied to produce a simulated light curve. 
The insets of Fig.~\ref{fig:gbm} show the results of this process for the Long and 
Short GRBs described above simulated for a timescale of $10^{-4}$ seconds and overall 
effective area of 100 square meters.

\subsection{Cross--Correlation technique and Monte-Carlo simulations}
 
\begin{figure}[h]
 \centering       
    \includegraphics[scale=0.5]{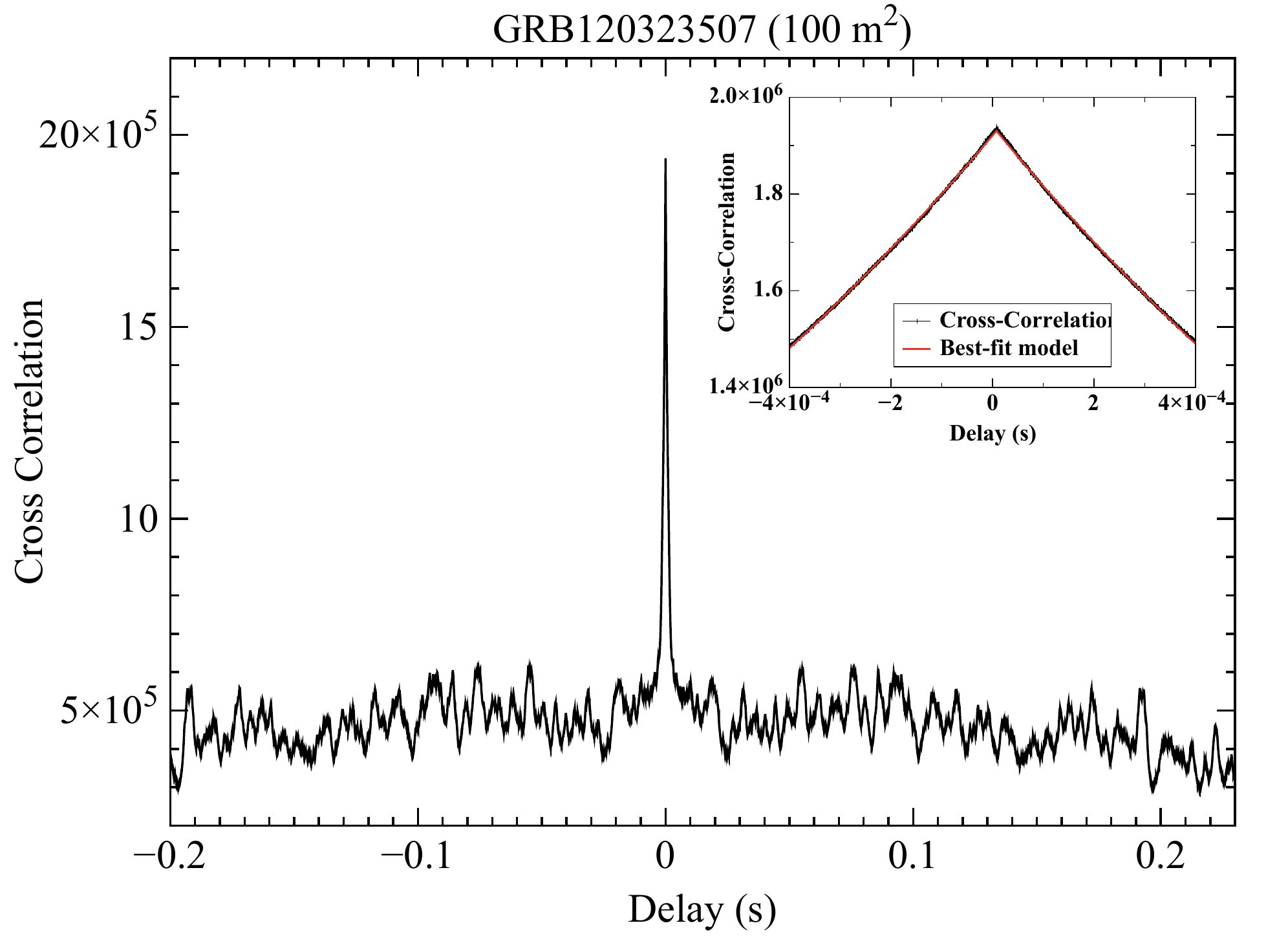}   
    \caption{Cross-correlation function obtained analysing simulated light curves obtained from a template generated starting from the Fermi GBM observations of the short GRB 120323507. See text for more details.}
    \label{fig:cross}
\end{figure}

Starting from the GRB light curves described above, we apply cross-correlation techniques to determine time delays between two signals. Fig.~\ref{fig:cross} shows an example of a cross-correlation function obtained by processing two GRB light curves simulated using the previously described template of the Short GRB observed by Fermi GBM (GRB120323507) that we rescaled to mimic a detector(s) with 100 square meters effective area. In order to extract the temporal information of the delay, we fitted a restricted region around the peak of the cross-correlation function with an \textit{ad hoc} model consisting of an asymmetric double exponential component (see inset in Fig.~\ref{fig:cross}). 
 
\begin{figure}[!h] 
   \centering      
    \includegraphics[scale=0.28]{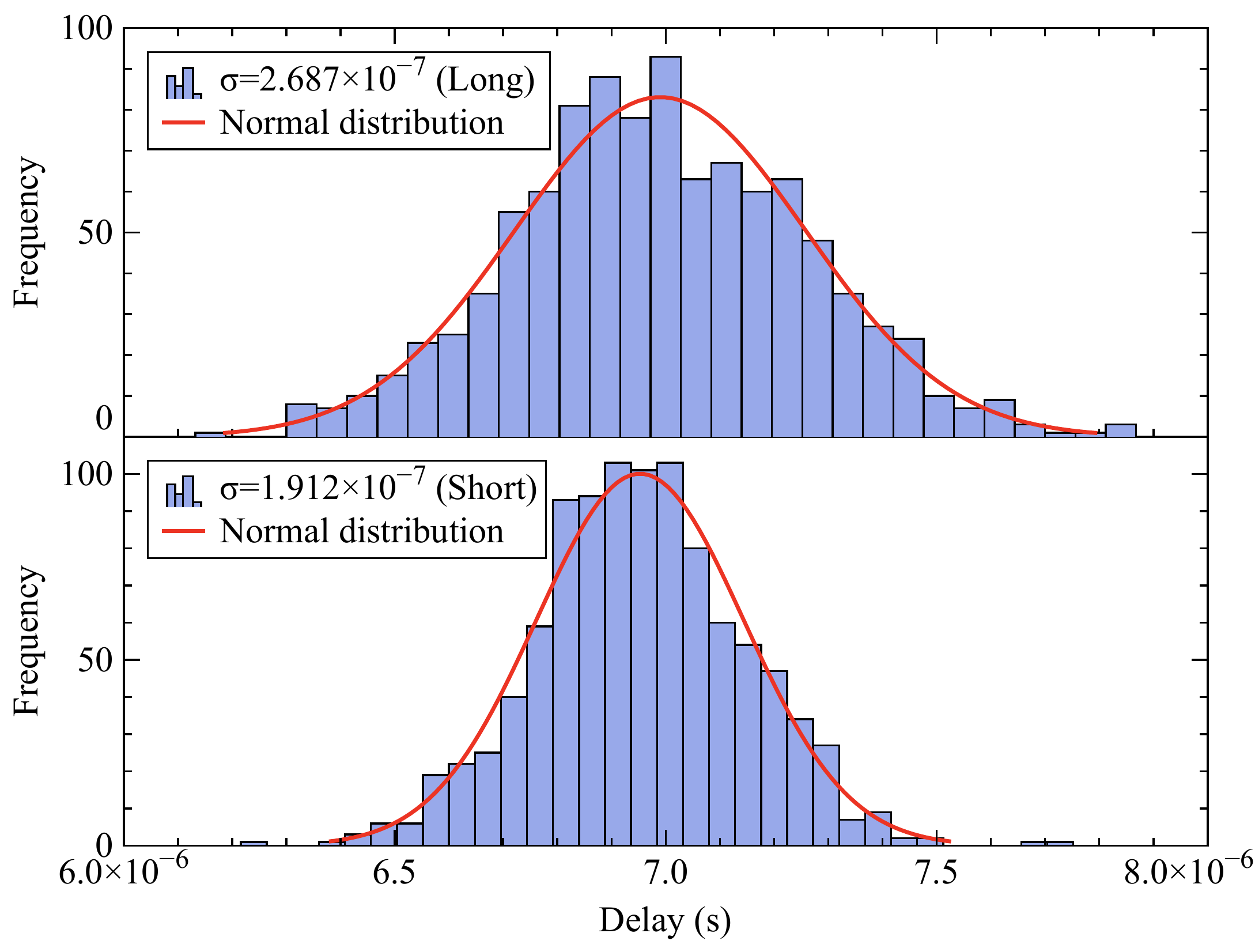}   
    \hspace{2px}
   \includegraphics[scale=0.28]{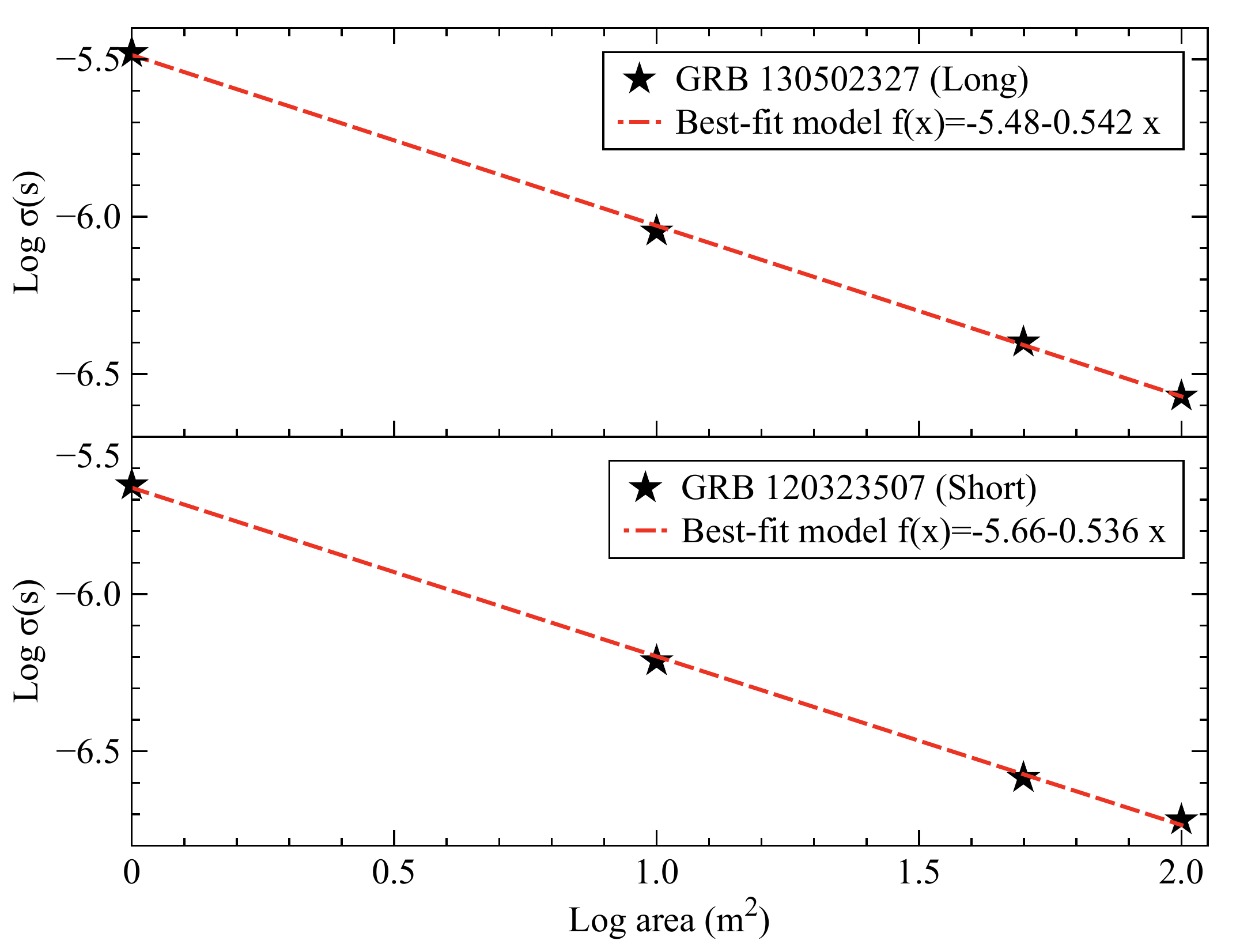}
    \caption{\textit{Upper panel:} Distribution of delays obtained applying cross-correlation techniques to pairs of simulated light curves of the Long (top) and the Short (bottom) Fermi GBM GRBs (see text for more details) rescaled for an effective collecting area of 100 square meters. Each distribution is the result of 1000 Monte-Carlo simulations. The overlaid red line represents the best-fit normal distribution to the data. \textit{Lower panel:} Dependence of the cross-correlation accuracy as a function of the effective area of the simulated instrument for the same Short and Long GRBs discussed in the upper panel. The red dashed line represents the best-fit model to the data.}
    \label{fig:dist}
\end{figure}

To investigate the accuracy achievable by the method, for each GRB and specific instrument effective area, we performed 1000 Monte-Carlo simulations in which two light curves generated by means of randomisation of the template are cross-correlated. For each cross-correlation function we then fitted the peak, extracting the delay between the light curves. From the overall distribution of delays we calculated its standard deviation which we interpret as a realistic estimate of the accuracy of the time delay measured with the cross-correlation method. The left panel of Fig.~\ref{fig:dist} shows the distribution of delays obtained from 1000 Monte-Carlo simulations performed for the Long (GRB130502327) and the Short (GRB120323507) GRBs assuming a total collecting area of 100 square meters. 
 
To proceed in the analysis of the technique we investigated the dependence of the cross-correlation accuracy, $E_{CC}$, as a function of the effective area of the instrument, which reflects the number of photons collected for the GRB. To do that, we performed 1000 Monte-Carlo simulations for two Short (one synthetic and one real) and two Long (one synthetic and one real) GRBs, simulating four different instrument collecting areas, i.e. 1, 10, 50 and 100 square meters, for a total of 16000 simulations. We note that each simulation performed on time scales of micro-seconds requires the creation of tens to hundreds millions of photons to be allocated in light curves with tens of millions of bins, which are then cross-correlated in pairs. 
The overall process involved a substantial computational effort, which required more than 6000 hours of CPU time in a multi-core (128 logical processors) server and several terabytes of storage.
 
From the simulations of the synthetic GRBs (in the band $\rm 50 \div 300 \, keV$) we obtained the following relations between the cross-correlation accuracy, $E_{CC}$, and the number of photons in the light curves $N_{ph}$:
$E_{CC\, {\rm Long}} = 0.014 \mu{\rm s}\times  (3.45\times 10^{8})^{0.634}\times N_{ph}^{-0.634}$ for the Long GRB and 
$E_{CC\, {\rm Short}} = 0.014 \mu{\rm s}\times  (6.1\times 10^{8})^{0.609}\times N_{ph}^{-0.609}$ for the Short GRB. 
 
From the simulations of the \textit{real} GRBs observed with Fermi GBM (in the band $\rm 50 \div 300 \, keV$) we obtained the following results (see also the lower panel of Fig.~\ref{fig:dist}):
$E_{CC\, {\rm Long}} = 0.27\mu{\rm s}\times  (2.83\times 10^{8})^{0.542}\times N_{ph}^{-0.542}$ for the Long GRB and 
$E_{CC\, {\rm Short}} = 0.19 \mu{\rm s}\times  (2.36\times 10^{7})^{0.536}\times N_{ph}^{-0.536}$ for the Short GRB. 

We can express these last relations in terms of GRB fluences $F$ and overall effective area of the detectors, $A$:
\begin{equation}
\label{eq:ecclong}
E_{CC\, {\rm Long}} = 0.27 \, \mu{\rm s}\times \left[ \left( \frac{F}{10^{-4}\,{\rm erg\, cm^{-2}}} \right) \left( \frac{A}{10^{2}\,{\rm m^{2}}} \right) \right]^{-0.542}
\end{equation} 
\begin{equation}
\label{eq:eccshort}
E_{CC\, {\rm Short}} = 0.19 \, \mu{\rm s}\times \left[ \left( \frac{F}{10^{-5}\,{\rm erg\, cm^{-2}}} \right) \left( \frac{A}{10^{2}\,{\rm m^{2}}} \right) \right]^{-0.536}
\end{equation} 
As expected, the cross correlation accuracy $E_{CC}$ scales roughly as the inverse square root of the GRB fluence $F$, and detector effective area $A$.
This shows that delays as small as a few $\mu{\rm s}$ can be detected with an effective area of $\sim 1 \, {\rm m}^2$.  


\section{\gq\ localisation capabilities}

\gq\ is designed to provide prompt (within seconds/minutes),
arcminute-to-(sub)-arcsecond localisations of bright hard X-ray transients. This
is the key to enable the search for faint optical transients
associated with the GWEs and GRBs, because their brightness quickly
fades after the event. In the \gq\ concept, localisation is achieved
by exploiting the delay between the transient's photon arrival times
at different detectors, separated by hundreds/thousands km. Delays are
measured by cross-correlating the source signals detected by different
instruments. 

The working principle of \gq\ can be easily understood by considering the
analogy with radio interferometry. 

In the case of radio interferometry obtained with $N$ observing radio telescopes
with average spatial separation $d$, the theoretical spatial resolution of the interferometric array 
results from the combination of $N_{\rm tot} = N \times (N-1)/2$ statistically {\it dependent} pairs of interferometers, 
each having an angular resolution capability of 
\begin{equation}
\label{eq:res1}
\sigma_{\theta, \; {\rm i}} \sim f(\alpha;\delta)_{\rm i} \times \sigma_{\phi \; i}  \times (\lambda/d),
\end{equation}
where $f(\alpha;\delta)_{\rm i} \mathcal{O}\!(1)$ is a function that depends on the position of the source in the 
sky ($\alpha$ and $\delta$ are the right ascension and declination, respectively) with respect to the orientation 
of the vector connecting the pair of antennas of the $\rm i_{th}$ interferometer, $\sigma_{\phi \; i}$ is the uncertainty in the 
phase differences measurable by each pair of antennas, $\lambda$ is the
wavelength of the observation and ${\rm i} = 1, ..., N$. 
It is important to note that the number of statistically independent pairs is $N_{\rm ind} = N -1$.
In practice, however, it is useful to consider the whole set of  $N_{\rm tot}$ equations to minimise the {\it a priori} unknown systematic effect on
one or more radio telescopes.
This system of $N_{\rm tot}$ equations can be solved for the 2 unknowns $\alpha$ and $\delta$ giving a statistical accuracy
of 
\begin{equation}
\label{eq:res2}
\sigma_\alpha \sim \sigma_\delta \sim g(\alpha;\delta) \times \sigma_{\phi} \times(\lambda/d)/ \sqrt{N_{\rm ind} - 2},
\end{equation} 
where $g(\alpha;\delta) \mathcal{O}\!(1)$ and $\sigma_{\phi}$ are suitably weighted averages of $f(\alpha;\delta)_{\rm i}$ and $\sigma_{\phi \; i}$, respectively.
The factor $\sigma_{\phi} \times \lambda$ represents the accuracy of the determination of the phases of the ratio signal.

In the case of \gq\ we can imagine that, because of the intrinsic variability of the signal of the transient sources, we are able 
to determine the analog of the factor $\sigma_{\phi} \times \lambda$ by cross-correlating the signal recorded by each pair of detectors
of the \gq\ constellation and determining the cross-correlation delay $\Delta t_i$. 
Indeed, since $\lambda \nu = c$, and $\phi = \int \nu dt \sim \nu \Delta t$ for short signals
(where $c$ is the speed of light and $\nu$ is its frequency), 
$\sigma_{\phi}  \times \lambda = \nu \sigma_{\Delta t} \lambda = c \sigma_{\Delta t}$, 
where $\sigma_{\Delta t}$ is a suitably weighted average (over the whole ensemble of detectors) 
of the accuracy in the determination of $\Delta t_i$.
Therefore, the accuracy in the source position obtainable with a constellation of $N$ satellites is
\begin{equation}
\label{eq:res3}
\sigma_\alpha \sim \sigma_\delta \sim g(\alpha;\delta) (c/d)\sigma_{\Delta t} / \sqrt{N - 3}.
\end{equation}  
Finally, we have to add in quadrature all the statistical errors in the determination of $\sigma_{\Delta t}$.
In particular we have:
\begin{equation}
\label{eq:dtacc}
\sigma_{\Delta t} = \sqrt{E_{\rm CC}^2+E_{\rm POS}^2+ E_{\rm time}^2}
\end{equation}
where 
$E_{\rm CC}$ is cross-correlation accuracy between the light curves recorded by two detectors,
$E_{\rm POS}$ is the error induced by the uncertainty in the spatial localisation of the detectors, and $E_{\rm time}$ is
the error in the absolute time reconstruction.
For large $N$, we adopt the reasonable value $g(\alpha;\delta) \sim 1$ and $N - 3 \sim N$,
$\sigma_\alpha \sim \sigma_\delta = \sigma_\theta$, where $\sigma_\theta$ is the positional accuracy (PA hereinafter):
\begin{equation}
\label{eq:res4}
\sigma_\theta \sim \frac{c}{d\sqrt{N}} \sqrt{E_{\rm CC}^2+E_{\rm POS}^2+ E_{\rm time}^2}.
\end{equation}

The absolute time and position reconstruction provided by commercial GPS systems are of the
order of 10-30 nanoseconds and $\sim10$ meters (corresponding again
to a few of tens nanoseconds). Moreover, we note that uncertainties in the times coming from the detection process must be taken into account. However, the intrinsic detection process and front end electronics readout can achieve sub-to-several nanoseconds accuracies and with careful design of the digital electronics, and a few nanoseconds timing can be achieved with heritage electronics. This leaves the error in the time delay
inferred from the cross-correlation analysis to be most likely the largest term within the time delay
uncertainty. 

Adopting $N_{100} = 100$ satellites for the constellation, 
$d_{\rm 3000 \; km} = 3 \times 10^8$ cm,
$E_{\rm CC \;10 \mu s} = 10^{-5}  \;  >> E_{\rm POS} >> E_{\rm time}$ 
we have
\begin{equation}
\label{eq:res5}
\sigma_\theta \sim 20.6 \; d_{\rm 3000}^{-1}  N_{100}^{-1} E_{\rm CC \;10 \mu s} \; {\rm arcsecond}.
\end{equation}


The PA calculated above includes statistical errors
only. Systematic errors are likely to be important, but at the stage
of proof of concept we can conclude that localisation at the sub-arcminute level
is feasible with the above parameter settings.

\section{High energy Transient localisation in the Multi--messenger Era}
\label{sec:2}

As of today, the observatories dedicated to the search and study of hard
X-ray transients are the NASA {\it Swift} and {\it Fermi}, and the ESA
{\it INTEGRAL} satellites.  

{\it Swift} was launched in 2004 and it is equipped with the wide field
of view (FoV) Burst Alert Telescope (BAT) to localise transients and
the narrow field X-ray Telescope (XRT) and the Ultra-Violet and Optical Telescope (UVOT), 
high sensitivity telescopes for detailed
observations of the transient afterglows. BAT is a coded mask
instrument with FoV$\sim$1/6 of the full sky, and a
collecting area of about 0.5 m$^2$ \citep{Barthelmy2004}. It can provide GRB positions above
1 arcminute accuracy, depending on GRB strength and position in the
FoV. XRT is a Wolter-I X-ray telescope, with FoV$\sim$30 arcminute$^2$,
and collecting area $\sim$200 cm$^2$, that can provide positions with
arcsecond accuracy of sources down to fluxes $\sim10^{-14}$ ergs/cm$^2$/s.  
{\it Swift} has the unique capability to slew from its
original pointing position to the position of the transient in tens of
seconds/minutes, to study the transient with its narrow field
telescopes.

{\it INTEGRAL} was launched in 2002 and it is equipped with the wide
field of view IBIS camera, FoV$\sim$1000 deg$^2$ and collecting area
$\sim1$ m$^2$ \citep{Ubertini2003}. IBIS has a smaller FoV than BAT, but a better
sensitivity, allowing the detection of fainter transients with respect 
to BAT. In addition to IBIS, the anti-coincidence scintillators 
of SPI, the high energy spectrometer, can be used 
as an all sky monitor to detect GRBs, with
basically no independent localisation capability, but very useful as a point in the Interplanetary Network.

{\it Fermi} was launched in 2008 and carries the GBM experiment,
consisting of 12 NaI and 2 BGO scintillators, each with about 120 cm$^2$
of collecting area \citep{Meegan09}. The GBM can provide GRB
positions with accuracies of several degrees in the best cases.

{\it Swift}, {\it INTEGRAL} and {\it Fermi} are working nominally after more than 12, 14
and 9 years from the launch respectively, providing $\sim$arcminute
positions ({\it Swift}, {\it INTEGRAL}) or tens of degrees positions ({\it Fermi}) over a
large fraction of the sky. Their predicted lifetimes would extend the
missions through the 2020's, but ageing of the equipment is ongoing and it is unknown how long they will survive
after 2020. This time window is crucial for two main
reasons:

\begin{enumerate}

\item 
The Advanced LIGO/Virgo detectors will reach their final sensitivity and best localisation capability for GWE in a few years. KAGRA joined the network at the beginning of 2020. However, a fifth interferometer, LIGO-India, will be required in the network (expected in 2025) to provide positions for a large fraction of GWE with accuracy smaller than 10 degrees.
On the other hand, the improved sensitivity will
increase the distance at which an event can be observed, to several Gpc
for BH-BH events and hundreds of Mpc for NS-NS events, thus increasing
the cosmic volume. The number of optical transients in such huge
volumes is from many tens to several hundreds, making it difficult to
identify the one associated with the GWE. The number of high-energy
transients in the same volume is much smaller, greatly helping the
identification. It is instructive to consider the first identification
of an electromagnetic transient with a GWE which occurred on August
17 2017. 
The {\it Fermi} GBM observed a gamma-ray burst within a few seconds of the GW detection.
The combined LIGO/Virgo error-box was the order
of 30 deg$^2$ \citep{GW170817}. However the LIGO/Virgo detection
indicated a very close event ($\sim$40 Mpc) greatly limiting the number
of target galaxies. An optical transient from one of these nearby
galaxies was soon discovered. There were thus two key elements that
allowed the discovery and localisation of the optical transient associated 
to the GWE: a) the prompt $\gamma$-ray detection from the {\it Fermi} GBM (and the Interplanetary Network triangulation with  
{\it INTEGRAL}), and b) the relatively limited volume that had to be searched.  
For fainter events, farther away, such those that will likely be provided by ground-based interferometers during the 2020s, the volume to be searched will be much larger. The third observing run of LIGO and Virgo already revealed events more distant than GW170817 for which a well-localised high-energy counterpart becomes crucial to detect the multi-wavelength signal and identify the host galaxy. 
The third generation of gravitational wave detectors is expected after 2030, e.g. the Einstein Telescope; at that time the localisation of possible GRB counterparts will be crucial \citep[see e.g.][]{Chan18} and \gq\ will be fundamental in this respect.

\item
By the end of the 2020s, ESA will launch its L2 mission {\it Athena},
carrying the most sensitive X-ray telescope and the highest energy resolution
detector (XIFU) ever built. Among the core science goals of {\it Athena} there
are spectroscopic observations of bright GRBs, used as 
light-beacons to X-ray the inter-galactic medium (IGM). These observations may lead to
the discovery and the characterisation of the bulk of the baryons in
the local Universe, in the form of a warm IGM (a few millions K),
through absorption line spectroscopy \citep[see e.g.][]{Fiore00}.  {\it Athena}
will also target high-z GRBs, to assess whether they are the final end
of elusive Pop-III stars (through the measurements of the abundance
pattern expected from the explosion of a star made only of pristine
gas). Indeed, very massive Pop-III stars are thought to collapse into proto-black holes. Subsequent accretion through a temporary disc could produce an energetic jet which, in turn, generates a burst of TeV neutrinos. This population of high energy neutrinos could be detected by the enhanced sensitivities of forthcoming detectors in the high-energy band such as AMANDA-II and IceCube \citep[][]{Schneider02}. This high redshift GRB population is intrinsically faint and therefore an ideal target for the unprecedented sensitivity of \gq\ . Moreover, because of the high redshift, quantum gravity time delays (if detectable) are significant in these systems. 

\end{enumerate} 

For these reasons several missions aimed at localising fast high
energy transients have been and will be proposed to NASA (Midex class)
and ESA (M class), to guarantee that the study of these elusive sources
can be operative and efficient during the next decades. \gq\ will
offer a fast-track and less expensive fundamental complement to these
missions, since it will be an all-sky monitor able to spot transient
events everywhere in the sky and to give a fast (within minutes) and 
precise (from below 1 deg to arcsecond, depending on the GRB flux and time
variability) localisation of the event. This is extremely important to
allow follow up observations of these events with the sensitive narrow field 
instruments of future complex and ambitious missions in all the bands 
of the electromagnetic spectrum (from radio to IR/Optical/UV and to X 
and gamma-rays).

The main parameters affecting the discovery space in this area are: 1)
number of events with good localisation; 2) quality of the
localisation; and 3) promptness of the localisation. \gq\ will ensure all
these three characteristics and will be fundamental to thoroughly study the
electromagnetic counterparts of GWE.

\section{Transients as tools to investigate the structure of space-time}
\label{delays}

\subsection{\gq\ Constellation as a single instrument of huge effective area}

Once the times of arrival (ToA) of the photons in each detector of the \gq\ constellation are corrected by the delays induced
by the  position of the GRB in the sky, as deduced from the optical identification of the counterpart, it is possible to add all the photons collected by the $N$ detectors of the constellation
to obtain a single light-curve equivalent to that of a single detector of effective area $A_{\rm tot} = N a$ where $a$ is the effective area
of each detector. In doing this an error in the ToA of each photon is introduced, because of the uncertainty in the position in the sky. However, since the optical counterpart will be known to within 1 arcsecond or below, the induced errors in the ToA are negligible.

\subsection{Is Vacuum a dispersive medium for photons?}
\label{delays}

As discussed in \S \, \ref{intro}, several theories proposed to describe quantum space-time predict a
discrete structure for space on small scales, $\ell_{\rm min} \sim \ell_{\rm P}$.
For a large class of these theories this space discretisation implies the onset of a dispersion relation
for photons, which could be related to the possible break or violation
of Lorentz invariance on such scales. Special Relativity postulates Lorentz invariance: all
observers measure the same speed of light in vacuum, independent of
photon energy, which is consistent with the idea that space is a three
dimensional continuum. On the other hand, if space is discrete on very
small scales, it is conceivable that light propagating in this
lattice exhibits a sort of dispersion relation, in which the speed of
photons depends on their energy.
These LIV models predict
a modification of the energy-momentum "dispersion" relation of the form
\begin{equation}
\label{eq:QG1}
E^2 = (pc)^2 + (mc^2)^2 + \Delta_{\rm QG}(E, p^2, M_{\rm QG})
\end{equation}
where $E$ is the energy of a particle of (rest) mass $m$ and momentum $p$, and
$M_{\rm QG} = \zeta M_{\rm P}$ is the mass at which quantum space-time
effects become relevant, where $\zeta \sim 1$, and
(since Special and General Relativity were thoroughly tested in the last century)
$\lim_{E/(M_{\rm QG}c^2) \rightarrow 0} \Delta_{\rm QG}(E, p^2, M_{\rm QG}) = 0$
\citep[see e.g.][]{Amelino00}.

In a very general way, the equation above can be used to determine the speed of a particle
(in particular a photon), given its energy. Moreover, when two photons of different energies, $E_2 - E_1 = \Delta E_{\rm PHOT}$, emitted at the same time,
travel over a distance $D_{\rm TRAV}$ (short with respect to the cosmic distance scale, {\it i.e.}
a distance over which the cosmic expansion can be neglected, see below), because of the dispersion relation above, they exhibit a
delay $\Delta t_{LIV}$.
It is possible to express this relation as a series expansion around
its limit value $\Delta t_{LIV} = 0$ (in line with what is discussed above we must have the following asymptotic condition:
$\lim_{E_{\rm PHOT}/(M_{\rm QG}c^2) \rightarrow 0} \Delta t_{LIV} = 0$) as:
\begin{equation}
\label{eq:dtLIV1}
\Delta t_{LIV} = \pm \xi \, (D_{\rm TRAV}/c) \, [\Delta E_{\rm PHOT}/(\zeta M_{\rm P}c^2)]^n
\end{equation}
where $\xi \sim 1$ is the coefficient of the first relevant term in the series expansion in the small parameter
$\Delta E_{\rm PHOT}/(M_{\rm QG}c^2)$, the sign $\pm$ takes into account the possibility
(predicted by different LIV theories) that higher energy photons are faster or slower than lower energy photons \citep[discussed as subluminal, $+1$, or
superluminal, $-1$, as in][]{Amelino09}.
Note that $\xi = 1$ in some specific LIV theories \citep[see e.g.][in particular their equation 13]{Camelia98,Amelino09}.
The index $n = 1 \; {\rm or} \; 2$ takes into account the order of the first non-zero term in the expansion.

When the distance traveled by the photons is comparable to the cosmic distance scale,
the term $D_{\rm TRAV}/c$ must be changed into $D_{\rm EXP}/c$ to take into account the effect of a particle propagating into an expanding Universe.
The comoving trajectory of a particle is obtained by writing its Hamiltonian in terms of the comoving momentum \citep{Jacob08}.
The distance traveled by the photons, in a general Friedman-Robertson-Walker Cosmology,
is determined by the different mass-energy components of the Universe.
These energy contents can be expressed in units of the critical energy density $\rho_{\rm crit} = 3 H_{0}^2/(8\pi G) = 8.62(12) \times 10^{-30}\; {\rm g/cm^3}$,
where $H_{0} = 67.74(46) \; {\rm km/s/Mpc}$ is the Hubble constant (see Planck Collaboration, 2015,
for the parameters and related uncertainties).
Considering the different dependencies on the cosmological scale factor $a$, it is possible to divide the energy components
of the Universe into: $\Omega_{\Lambda} = \rho_{\Lambda}/\rho_{\rm crit}$,  $\Omega_{\rm M} = \rho_{\rm Matter}/\rho_{\rm crit}$,
$\Omega_{\rm R} = \rho_{\rm Radiation}/\rho_{\rm crit}$, $\Omega_{\rm k} = 1 - (\Omega_{\Lambda} + \Omega_{\rm M} + \Omega_{\rm R})$.
With this notation it is possible to express the proper distance $D_{\rm P}$ at the present time (or comoving distance)
of an object located at redshift $z$ as:
\begin{equation}
\label{eq:properdist}
D_{\rm P} = \frac{c}{H_{0}} \int_0^z dz \frac{1}{\sqrt{f(\Omega,z)}},
\end{equation}
where
\begin{equation}
\label{eq:expansion}
f(\Omega,z) = (1+z)^{3(1+w)}\Omega_{\Lambda} + (1+z)^2\Omega_{\rm k}
+ (1+z)^3\Omega_{\rm M} + (1+z)^4\Omega_{\rm R},
\end{equation}
On the other hand, the term $D_{\rm EXP}$ has to take into account the fact that the proper distance varies as the universe expands.
Photons of different energies are affected by different delays along the path, so, because of cosmological expansion, a delay produced further back
in the path amounts to a larger delay on Earth. This effect of relativistic dilation introduces a factor of (1 + z) into the above integral \citep{Jacob08}.
\begin{equation}
\label{eq:traveldist}
D_{\rm EXP} = \frac{c}{H_{0}} \int_0^z dz \frac{(1+z)}{\sqrt{f(\Omega,z)}},
\end{equation}

In particular, in the so-called Lambda Cold Dark Matter Cosmology ($\Lambda$CDM) the following values are adopted (Planck Collaboration, 2015):
$H_{0} = 67.74(46)$ ${\rm km\,s^{-1}Mpc^{-1}}$,
$\Omega_{\rm k} = 0$, curvature $k=0$ that implies a flat Universe,  $\Omega_{\rm R} = 0$, radiation $= 0$ that implies a cold Universe,
$w = -1$, negative pressure Equation of State for the so-called Dark Energy that implies an accelerating Universe,
$\Omega_{\Lambda} = 0.6911(62)$ and $\Omega_{\rm Matter} = 0.3089(62)$.
With these values we have:
\begin{equation}
\label{eq:traveldist}
\frac{D_{\rm EXP}}{c} = \frac{1}{H_{0}} \int_0^z dz \frac{(1+z)}{\sqrt{\Omega_{\Lambda} +
+ (1+z)^3\Omega_{\rm Matter}}}.
\end{equation}
Adopting as a firm upper limit for the distance of any GRB the radius of the visible (after recombination) Universe
$D_{\rm P}/c \le R_{\rm V}/c = 1.4 \times 10^{18} \,{\rm s}$ (in the $\Lambda$CDM cosmology),
we find:
\begin{equation}
\label{eq:dtLIV2}
| \Delta t_{LIV} | \le 1.4 \times 10^{18} \xi \; [\Delta E_{\rm PHOT \, MeV}/ (\zeta \times 10^{21})]^n \, {\rm s}
\end{equation}
where $\Delta E_{\rm PHOT \, MeV} = \Delta E_{\rm PHOT}/ (1\; {\rm MeV})$.
This shows that first order effects ($n =1$) would result in potentially detectable delays, while second order effects
are so small that it would be impossible to detect them with this technique.

Therefore, it is possible to detect (or constrain) first order effects in space-time quantisation by detecting (or placing upper limits on)
time delays between light curves of GRBs in different energy bands.
Indeed
these quantum-space-time effects modifying the propagation of light are
extremely tiny, but they cumulate along the way.
GRBs are among the best candidates to detect the expected delays, since
i) the signal travels over cosmological distances; ii) the prompt spectrum covers more than three order
of magnitudes in energy; iii) fast variability of the light-curve is present at or below the one millisecond level
\citep[see e.g.][]{Camelia98}.
Such a detection could directly reveal, for the
first time, the deepest structure of quantum space--time by gauging its
structure in terms of a photon dispersion relation {\it in vacuo}.

To better quantify this possibility, we considered a broad band,
$5\; {\rm keV}-50\; {\rm MeV}$, covering a relevant fraction of the prompt emission of a typical
GRB and within the energy range covered by NaI and BGO scintillators.
Based on BATSE observations of GRB prompt spectra, the so-called {\it Band function},
an empirical function describing the photon energy distribution, has been developed \citep{Band93}:
\begin{equation}
\label{eq:band}
\frac{dN_{E}(E)}{dA\; dt} =  F \times \left\{
\begin{array}{lr}
\left( \frac{E}{E_{\rm B}} \right)^{\alpha} \exp\{-(\alpha - \beta)E/E_{\rm B}\}, & E \le E_{\rm B},\\
\left( \frac{E}{E_{\rm B}} \right)^{\beta} \exp\{-(\alpha - \beta) \}, & E \ge E_{\rm B}.
\end{array}
\right.
\end{equation}
where
$E$ is the photon energy,
$dN_{E}(E)/(dA\; dt)$ is the photon intensity
energy distribution in units of ${\rm photons/cm^{2}/s/keV}$,
$F$ is a normalisation constant in units of ${\rm photons/cm^{2}/s/keV}$,
$E_{\rm B}$ is the break energy, and $E_{\rm P} = [(2+\alpha)/(\alpha - \beta)] E_{\rm B}$, which is the peak energy.
For most GRBs: $\alpha \sim -1$, $\beta \sim -2.5$, $E_{\rm B} \sim 225\; {\rm keV}$ gives
$E_{\rm P} = 150\; {\rm keV}$.


As representative spectra of long and short bright GRBs, we considered Band functions with $\alpha = -1$, $\beta = -2.5 \div -2.0$
(proxies of soft and hard GRB spectra),
$E_{\rm B} = 225\; {\rm keV}$ lasting for $\Delta t = 25 \div 0.25\;{\rm s}$ respectively, having a photon flux in the band $50-300\; {\rm keV}$ of
\begin{equation}
\label{eq:norm}
\displaystyle\int_{50\,{\rm keV}}^{300\,{\rm keV}} \frac{dN_{E}(E)}{dA\; dt} dE =
\frac{dN_{50-300\; {\rm keV}}}{dA\,dt} = 8\; {\rm photons/cm^{2}/s}.
\end{equation}
We computed the total number of photons
detected in 8 contiguous energy bands
$\Delta E_{\rm E_i \div E_{i+1}}$ ($i = 1,...,8$)
in the interval considered above ($5\; {\rm keV}-50\; {\rm MeV}$),
adopting a cumulative effective area of $100$ square meters.

Moreover, we considered three values of the redshift, namely $z = 0.1,\, 1,\, 3$ for the upper extreme of the integral in equation (\ref{eq:traveldist}),
adopted $\xi =1$, $\zeta =1$, and $n=1$ in equation (\ref{eq:dtLIV1}), substituted $D_{\rm TRAV}$ of equation (\ref{eq:traveldist}) with $D_{\rm EXP}$ in (\ref{eq:dtLIV1}), and computed the delays expected for each value of
$z$ and $\Delta E_{\rm PHOT \; i} = \sqrt{\rm E_i \times E_{i+1}}$\footnote{
The choice of using the geometric average (instead of the average) to consider the delays induced by a first order LIV violation typical of the given energy band,
is done to take into account the fact that GRB spectra decrease as a power-law, and, therefore, the lower limit of the band is richer in photons. The use of the linear average
has the effect of slightly increasing the computed delays.}. The results are shown in Table \ref{ttt}.
\begin{table*}

\begin{center}
\title{Long GRB (GRB130502327) --  $\Delta t = 30$ s}
\vskip 0.2cm
\bf{
\begin{tabular}{crccccrrrrr}
\hline
\hline
\\
\multicolumn{1}{c}{${\rm Energy \; band}$}    & \multicolumn{1}{c}{$E_{\rm AVE}$} & \multicolumn{1}{c}{$N$} &
\multicolumn{1}{c}{$E_{CC}(N)$} &
\multicolumn{1}{c}{$N$}  &
\multicolumn{1}{c}{$E_{CC}(N)$} &
\multicolumn{4}{c}{${\rm \Delta T_{\rm LIV}\; (\xi = 1.0, \; \zeta = 1.0)}$} \\
 & & $(\beta = -2.5)$ & & $(\beta = -2.0)$ & & & & & \\
\multicolumn{1}{c}{${\rm MeV}$}    & \multicolumn{1}{c}{${\rm MeV}$} & \multicolumn{1}{c}{${\rm photons}$} &
\multicolumn{1}{c}{$\mu{\rm s}$} &
\multicolumn{1}{c}{${\rm photons}$}  &
\multicolumn{1}{c}{$\mu{\rm s}$} &
\multicolumn{1}{c}{$\mu{\rm s}$} & \multicolumn{1}{c}{$\mu{\rm s}$} &
\multicolumn{1}{c}{$\mu{\rm s}$} & \multicolumn{1}{c}{$\mu{\rm s}$} \\
 & & & & & &
\multicolumn{1}{c}{$z=0.1$} &
\multicolumn{1}{c}{$z=0.5$} & \multicolumn{1}{c}{$z=1.0$} & \multicolumn{1}{c}{$z=3.0$} \\
\\
$0.010 - 0.025$ & $0.0158$ & $2.98 \times 10^8$ & $0.26$ & $2.39 \times 10^8$ & $0.29$ & $0.06$ & $0.35$ & $0.72$ & $2.01$ \\
$0.025 - 0.050$ & $0.0353$ & $1.98 \times 10^8$ & $0.33$ & $1.66 \times 10^8$ & $0.36$ & $0.13$ & $0.72$ & $1.46$ & $4.10$ \\
$0.050 - 0.100$ & $0.0707$ & $1.56 \times 10^8$ & $0.37$ & $1.41 \times 10^8$ & $0.39$ & $0.27$ & $1.43$ & $2.93$ & $8.21$ \\
$0.100 - 0.300$ & $0.1732$ & $1.27 \times 10^8$ & $0.42$ & $1.42 \times 10^8$ & $0.39$ & $0.66$ & $3.51$ & $7.19$ & $20.10$ \\
$0.300 - 1.000$ & $0.5477$ & $2.92 \times 10^7$ & $0.92$ & $5.41 \times 10^7$ & $0.66$ & $2.09$ & $11.11$ & $22.72$ & $63.56$ \\
$1.000 - 2.000$ & $1.4142$ & $3.72 \times 10^6$ & $2.82$ & $1.16 \times 10^7$ & $1.52$ & $5.40$ & $28.68$ & $58.67$ & $164.12$ \\
$2.000 - 5.000$ & $3.1623$ & $1.52 \times 10^6$ & $4.59$ & $6.96 \times 10^6$ & $2.01$ & $12.07$ & $64.12$ & $131.19$ & $367.00$ \\
$5.000 - 50.00$ & $15.8114$ & $4.98 \times 10^5$ & $8.40$ & $4.17 \times 10^6$ & $2.67$ & $60.35$ & $320.62$ & $656.00$ & $1834.98$ \\
\\
\hline
\hline
\end{tabular}
}
\end{center}
\begin{center}
\title{Short GRB (GRB120323507) -- $\Delta t = 0.4$ s}
\vskip 0.2cm
\begin{tabular}{crccccrrrrr}
\hline
\hline
\\
\multicolumn{1}{c}{${\rm Energy \; band}$}    & \multicolumn{1}{c}{$E_{\rm AVE}$} & \multicolumn{1}{c}{$N$} &
\multicolumn{1}{c}{$E_{CC}(N)$} &
\multicolumn{1}{c}{$N$}  &
\multicolumn{1}{c}{$E_{CC}(N)$} &
\multicolumn{4}{c}{${\rm \Delta T_{\rm LIV}\; (\xi = 1.0, \; \zeta = 1.0)}$} \\
 & & $(\beta = -2.5)$ & & $(\beta = -2.0)$ & & & & & \\
\multicolumn{1}{c}{${\rm MeV}$}    & \multicolumn{1}{c}{${\rm MeV}$} & \multicolumn{1}{c}{${\rm photons}$} &
\multicolumn{1}{c}{$\mu{\rm s}$} &
\multicolumn{1}{c}{${\rm photons}$}  &
\multicolumn{1}{c}{$\mu{\rm s}$} &
\multicolumn{1}{c}{$\mu{\rm s}$} & \multicolumn{1}{c}{$\mu{\rm s}$} &
\multicolumn{1}{c}{$\mu{\rm s}$} & \multicolumn{1}{c}{$\mu{\rm s}$} \\
 & & & & & &
\multicolumn{1}{c}{$z=0.1$} &
\multicolumn{1}{c}{$z=0.5$} & \multicolumn{1}{c}{$z=1.0$} & \multicolumn{1}{c}{$z=3.0$} \\
\\
$0.010 - 0.025$ & $0.0158$ &2.48 $ \times 10^7$ & $0.18$ & $1.99 \times 10^7$ & $0.21$ & $0.06$ & $0.35$ & $0.72$ & $2.01$ \\
$0.025 - 0.050$ & $0.0353$ & $1.65 \times 10^7$ & $0.23$ & $1.38 \times 10^7$ & $0.25$ & $0.13$ & $0.72$ & $1.46$ & $4.10$ \\
$0.050 - 0.100$ & $0.0707$ & $1.30 \times 10^7$ & $0.26$ & $1.18 \times 10^7$ & $0.27$ & $0.27$ & $1.43$ & $2.93$ & $8.21$ \\
$0.100 - 0.300$ & $0.1732$ & $1.06 \times 10^7$ & $0.29$ & $1.18 \times 10^7$ & $0.27$ & $0.66$ & $3.51$ & $7.19$ & $20.10$ \\
$0.300 - 1.000$ & $0.5477$ & $2.44 \times 10^6$ & $0.67$ & $4.51 \times 10^6$ & $0.46$ & $2.09$ & $11.11$ & $22.72$ & $63.56$ \\
$1.000 - 2.000$ & $1.4142$ & $3.10 \times 10^5$ & $1.94$ & $9.67 \times 10^5$ & $1.05$ & $5.40$ & $28.68$ & $58.67$ & $164.12$ \\
$2.000 - 5.000$ & $3.1623$ & $1.27 \times 10^5$ & $3.12$ & $5.80\times 10^5$ & $1.38$ & $12.07$ & $64.12$ & $131.19$ & $367.00$ \\
$5.000 - 50.00$ & $15.8114$ & $4.16\times 10^4$ & $5.69$ & $3.48 \times 10^5$ & $1.82$ & $60.35$ & $320.62$ & $656.00$ & $1834.98$ \\
\\
\hline
\hline
\end{tabular}
\end{center}
\caption{Photon fluence and expected delays induced by LIV for bright Long and Short GRBs observed with a detector of effective collecting area of $100\; {\rm m}^2$.
The GRB is described by a Band function
with $\alpha = -1$, $\beta = -2.5\div-2$, $E_{\rm B} = 225\; {\rm keV}$. The proper distance traveled by the photons has been computed for each redshift
adopting a $\Lambda$CDM cosmology with $\Omega_{\Lambda} = 0.6911$ and $\Omega_{\rm Matter} = 0.3089$. This implies the following proper
distances at the present time:
$D_{\rm EXP} = 453.9\, {\rm Mpc}$ for $z=0.1$,
$D_{\rm EXP} = 2411.4\, {\rm Mpc}$ for $z=0.5$,
$D_{\rm EXP} = 4933.6\, {\rm Mpc}$ for $z=1.0$, $D_{\rm EXP} = 13801.2\, {\rm Mpc}$ for $z=3.0$.
Adopting $\xi =1$, $\zeta =1$, and $n=1$ we found:
$| \Delta t_{LIV} | = 3.8168\, {\rm \mu s} \times \Delta E_{\rm PHOT}/ (1\; {\rm MeV})$ for $z=0.1$,
$| \Delta t_{LIV} | = 20.2775\, {\rm \mu s} \times \Delta E_{\rm PHOT}/ (1\; {\rm MeV})$ for $z=0.5$,
$| \Delta t_{LIV} | = 41.4863\, {\rm \mu s} \times \Delta E_{\rm PHOT}/ (1\; {\rm MeV})$ for $z=1.0$, $| \Delta t_{LIV} | = 116.0544\, {\rm \mu s} \times \Delta E_{\rm PHOT}/ (1\; {\rm MeV})$ for $z=3.0$. $\Delta E_{\rm PHOT} = E_{\rm AVE} = \sqrt{\rm E_{max} \times E_{min}}$ (see text).}
\label{ttt}
\end{table*}


Recent Fermi LAT detections of short GRBs at GeV energies have put
constraints on $\Delta t$ , and thus on $M_{QG}$ knowing $D(z)$. The
best limit so far was obtained by \citet{Abdo09b} using the short
GRB GRB090510. They find $\Delta t/\Delta E \ls 30$ms/GeV, which puts
$M_{QG} \sim M_{\rm Planck}$, at the distance of this GRB (z=0.9). This
limit, however, is obtained by assuming that a single observed 31 GeV
photon was emitted simultaneously to the other $\sim$GeV photons of the
burst, that lasted for $\sim0.2$s.

Indeed, a significant class of theories of Quantum Gravity describing the space--time structure down to the Planck scale predict a dispersion law for the propagation of photons {\it in vacuo} that depends linearly on the ratio between the photon energy and the Planck energy. The delays induced by this relation of light dispersion depend linearly on the space travelled and are tiny, being, as shown in Table \ref{ttt}, in the microsecond range, for photons that travelled for (few) billion years. 
Gamma--ray Bursts are ideal targets to test, robustly, this prediction because the prompt gamma--ray emission extends, in a detectable way, over more than six orders of magnitude in energy (from keV to ten(s) of GeV) and are among the most distant objects ever detected (their maximum redshift measured to date is just above 9). Intrinsic spectral delays due to unknown characteristics of the emission process in different energy bands could easily dominate the delays observable between different spectral components, but these effects can be disentangled by i) having a sufficient number of photons in sufficiently narrow energy bands, as the emission process is the same within a narrow band; ii) having a sufficiently rich sample of Gamma--ray Bursts at different redshifts, since the delays induced by a dispersion law for the propagation of photons {\it in vacuo} scale almost linearly (with a weak dependence on the details of the particular cosmology adopted) with redshift. This double linear dependence, in energy and redshift, is the characteristic signature of a Quantum Gravity effect.

Recently, \citet{Xu2016a,Xu2016b} and \citet{Amelino17a}
found {\it in--vacuo}--dispersion--like spectral lags in GRBs seen by Fermi
LAT. The magnitude of these effects is of the order of tens
$M_{\rm Planck}$, much bigger than the limit reported above obtained on
GRB090510. The effects are present when considering photons with
rest--frame energies higher than 40 GeV \citep{Xu2016a,Xu2016b}, or 5 GeV
\citep{Amelino17a}.
If this is the case, the predicted delays are one order of magnitude larger than those presented in Table \ref{ttt}.
\section{Astrophysical science with \gq\ }

Taking advantage of its huge effective area and the unprecedented timing capabilities, \gq\ 's science goals constitute {\it per se} an important milestone of astrophysical research; in the following we just list the main objectives of this ancillary science:

\begin{itemize}
\item To produce a catalogue of $7,000 \div 10,000$ GRBs with well determined positions in the sky (between 
$1^{\circ}$ and few arcsecond, depending on the flux and temporal variability of the GRB). Indeed, the expected number of GRBs in the whole sky is 2-3 per day and we plan to have a lifetime for this mission of at least ten years (note that single satellite failure will not be a problem since these can be easily replaced with high-performance newer versions). With the temporal triangulation technique previously described, position determination would be possible within minutes of the prompt event, allowing a search for its counterpart in other wavelengths. {\it Swift}-BAT allows localisation of GRBs occurring in its field of view with an accuracy of a few arcminutes (FoV of 1/6 of the sky), with the possibility for all of them to get an X-ray localization with XRT, and for some of them to get a subsequent optical localisation (with the UVOT) resulting in the determination of the redshift of their host galaxies. Similarly, the fast and precise GRB localisation offered by \gq\, solely from gamma-ray observations, will allow the determination of the optical counterpart and redshift for most of the long GRBs and for the short GRBs for which an optical counterpart can be detected. Since the counterpart of the furthest GRBs may fall in the IR band because of the high redshift, once a precise localisation of the source is found, it can be effectively searched thanks to the synergy with e.g. the James Webb Space Telescope (operating in the IR band); this will allow the detection of GRBs with $z>10$ (the actual record is just above $z=9$, \citep{Cucchiara_2011}), opening a brand new window for high-redshift cosmology. Moreover, if a dedicated mission such as {\it THESEUS} (selected for a possible ESA M5 mission) is approved by ESA, it would be totally synergetic with \gq\ since follow up observations of both soft X-ray localisations (obtained by {\it THESEUS} itself) and harder X-ray (or soft gamma-ray) localisation obtained with \gq would be possible. 

\item Given the huge effective area, \gq\ will be the ultimate experiment for prompt GRB physics. In this context we plan to produce a catalogue of GRB dynamic spectra over more than three orders of magnitude in energy (from 20 keV to 10 MeV) with unprecedented statistics and moderate energy resolution. Again, the combination of huge effective area and high time resolution will allow us to have enough photons in the high-energy band to follow spectral evolution of the prompt emission on short timescales. This is particularly important to shed light on the complex and poorly studied details of the fireball model and the mechanism through which ultra-relativistic colliding shocks release the huge amount of gamma-ray photons observed in the GRB's inner engine. GRBs are thought to be produced by the collapse of massive stars and/or by the coalescence of two compact objects. Their main observational characteristics are the huge luminosity and fast variability, often as short as one millisecond. These characteristics soon led to the development of the fireball model, i.e. a relativistic bulk flow where shocks efficiently accelerate particles. The cooling of the ultra-relativistic particles then produces the observed X-ray and gamma-ray emission. While successful in explaining GRB observations, the fireball model implies a thick photosphere, hampering direct observations of the hidden inner engine that accelerates the bulk flow. We are then left in the frustrating situation where we regularly observe the most powerful accelerators in the Universe, but we are kept in the dark over their operation. GRB fast variability is potentially the key to reveal the nature of their inner engines.  Early numerical simulations \citep[see e.g.][]{Kobayashi97,Ramirez00}, as well as modern hydro-dynamical simulations \citep{Morsony10}, and analytic studies \citep[see e.g.][]{Nakar02} suggest that the GRB light-curve reproduces the activity of the inner engine. GRB light-curves have been investigated in some detail down to 1 msec or slightly lower \citep{Walker00,MacLachlan13}. Sub--msec timescales are basically unknown, as little known as the real duration of the prompt event. Furthermore, it is still unclear how many shells are ejected from the central engine, what is the frequency of ejection and what their lengths are. Pushing GRB timing capabilities by more than three orders of magnitude should help in answering at least some of these questions.

\item To add polarimetric information on the sample of GRBs detected. \citet{McConnell_1996} proposed to measure the linear polarisation of GRBs by comparing the asymmetry in the rate of counts of the delayed component of photons Compton-backscattered by the Earth's atmosphere as observed by different {\it BATSE} detectors. This technique might be applied to data collected by \gq\ by comparing the photons detected by different satellites at different directions with respect to the Earth and by exploiting the timing capabilities of its instruments; in this case the method will be much more effective. Polarisation will provide other valuable information of extreme interest for the fireball model.
Results from POLAR, a dedicated GRB polarimeter onboard China's Tiangong--2 space laboratory, suggest that the gamma-ray emission is at most polarized at a relatively low level. However, the results also show intra--pulse evolution of the polarization angle. This indicates that the low polarization could be due to a variation of the polarization angle during the GRB \citep[][]{Zhang2019}. Given the superb temporal resolution and huge effective area of \gq\, this possibility will be thoroughly explored.

\item To scrutinise the whole sky for X and gamma-ray transients of very short duration. Despite its lack of imaging capabilities, \gq\ will benefit from the fact that background is relatively low at energies above few tens of keV. The huge effective area will guarantee an unprecedented sensitivity allowing the detection (signal-to-noise ratio $> 1$) of transient phenomena at the shortest timescales, mitigating the effects of the quantum-detection process that are blinding our sensitivity when the number of photons detected is small. There might exist a large class of fast transients that have remained undiscovered up to now because of the small fluence associated with their short time duration. In the radio band this has been the case of the recently discovered Fast Radio Bursts \citep[FRBs, see][as a review]{Lorimer_2018}. Indeed, some theories predict, and observations have now confirmed, a high energy counterpart of these compelling phenomena and \gq\ is the right instrument for searching these counterparts.
In particular, high energy counterparts are predicted in the context of Quantum Gravity \citep{Barrau14}.
In the same context it is possible that black holes hide a core of Planckian density, sustained by quantum gravitational pressure.  As a black hole evaporates, the core remembers the initial mass and the final explosion occurs at macroscopic scale. Under several rough assumptions, it is possible to estimate that several short gamma--ray events per day, at energies around 10 MeV, with isotropic distribution, can be expected coming from a region of a few hundred light years around us.
Further predictions can be done, in particular, to show that the wavelength of these signals should depend on the size of the black hole at the moment of the explosion \citep{Barrau14}.

\item To monitor all kinds of high-energy transients, both galactic and extra-galactic events, such as the flaring activity of magnetars, and outbursts of black hole and neutron star transients. The monitoring of the high-energy sky has been very important in the last years in the discovery of new events and/or peculiar behaviours as well as for a detailed characterisation of known sources. \gq\ will perform as a large area all-sky monitor, with good temporal and moderate energy resolution, able to add important information for the full understanding and the thorough study of high-energy transients, whose behaviour may lead to important advances in fundamental physics regarding strong gravity and extremely high-density matter.

\item To monitor the onset of Tidal Disruption Events (TDE, hereafter) with fast variability. 
Tidal disruption events \citep{Rees88} are generally very luminous (often above Eddington) in the soft X-ray band, with an X-ray spectrum usually dominated by a thermal component at a few keV \citep{Holoien16}. However, a sub-class of TDEs, called ``jetted TDEs'' are characterised by a much harder non-thermal spectrum extending up to the gamma-ray band \citep[see the prototypical case of Swift J16644;][]{Bloom11}. They are a fundamental tool in the study of the ``onset'' of AGN-like activity in otherwise quiescent black holes. Since most of the emission arises close to the black hole, they can be used to study relativistic phenomena such as precession induced by the black hole spin \citep{pasham19}. Also, they can serve as an important probe of hidden, sub-pc black hole binaries that are in the process of merging and are thus progenitors of LISA events \citep{Vigneron18}. Finally, TDEs also produce dim, but potentially detectable gravitational wave emission \citep{Kobayashi04} and might thus be important electromagnetic counterparts to a sub-class of gravitational wave sources.

\item To perform high-quality timing studies of known high-energy pulsators. The most interesting sector of this population contains the millisecond pulsars \citep[accreting and/or transitional and/or rotationally powered, see e.g.][]{DiSalvo_2018} and the enigmatic gamma-ray pulsars. Millisecond pulsars often display (transient) X-ray and gamma-ray emission whose properties are not completely understood yet. This emission may be caused by intra-binary shocks in the pulsar emission (consisting of both radiation and high-energy particles) with a wind of matter from the companion star. In this case, a modulation of the X- and gamma-ray emission with the orbital period is expected and may be searched for with \gq. Also, the orbital period evolution of these systems is very important to address in order to investigate their formation history and their connection with Low Mass X-ray Binaries, as envisaged by the recycling scenario. Orbital evolution may also be studied in high inclination X-ray binary systems (containing black holes or neutron stars) where periodic signatures (such as dips and/or eclipses) are observed.
Despite the lack of imaging capabilities and no possibility of background rejection, \gq\ is capable of detecting any (quasi-)periodic signal for which the period is known thanks to folding techniques coupled with a huge collecting area. This makes this instrument an ideal tool to perform timing studies of any kind of high-energy (quasi-)periodic signal. 

\end{itemize}


\section{Detector description}
The key requirements for a \gq\ detector are:  
\begin{itemize}
\item Overall effective area of the order of $100 \, {\rm m}^2$. This is obtained with a fleet consisting of tens/hundreds/thousands
of small/micro/nano satellites each hosting a detector of effective area ranging from $\sim 1\, {\rm m}^2$ to $\sim 100\, {\rm cm}^2$.
\item Capability of recording each photon (event) of the signal (no pile-up). 
\item Temporal resolution in the 10--100 nanoseconds range 
\item
Wide energy band from a few keV to several MeV. 
\item
Moderate energy resolution: $\Delta E / E \leq 0.2$ throughout the entire energy band.
\item
Wide field of view ($\sim$ steradians). 
\item
Robust assembly suitable for space environment.
\item
Simple design to allow for mass production.
\end{itemize}
A class of X/gamma detectors, widely used in countless space
experiments, that is continuously renewed thanks to evolving technology, is based on the use of scintillators coupled
to suitable photodetectors and electronics. Nowadays, inorganic
scintillator materials like Lanthanum Bromide (LaBr3:Ce), GAGG
(Gadolinium Aluminium Gallium Garnet) or similar, combine high
scintillation light emission with fast response (tens of nanoseconds),
and high efficiency. We therefore have, today, a certain number of materials whose characteristics allow, when
combined with a fast and efficient photodetector, the fulfillment of the
\gq\ project requirements. 
The criteria for the choice of scintillator
can then take into account parameters like intrinsic low background of
the material, low hygroscopicity, low cost, and low radiation damage.  A fast
photodetector for the readout of the scintillation light can be a
Photomultiplier (PMT) or solid state Silicon-PMT (Si-PM), both devices
having a response to a light pulse than can be contained in few
nanoseconds. Alternatively, Silicon Drift Detectors (SDDs) can be used to
read out the scintillation light with timing capabilities of the order
of tens of nanoseconds. Despite their relatively lower response to
light pulses, SDDs have several advantages with respect to Si-PM,
namely their greater robustness against radiation environment and
higher efficiency (90\% vs. 20-30\%).  Both kinds of devices, when
optically coupled to the above mentioned scintillators, allow efficient
detection of X-rays down to $\sim 10$ keV and even below. The criteria for the choice of the photodetector can take into
account the dimensions and robustness of the device, its ageing in the space radiation environment, and the availability for mass
production.  The architecture of each \gq\ detector sub-unit is modular, with modules 
of few cm$^2$ of geometric area each.  
The whole detector is then assembled to the
necessary size by adding modules, which will also ease the processing of
intense impulsive events by reducing the pile-up of signals in any given
module. 

\section{Conclusion: \gq\ mission concept}

The planning of the ESA  Science Programme Voyage  2050 relies on the public discussion of 
open scientific questions of paramount importance for an advance in our understanding of the Laws 
of Nature, that can be addressed by a scientific space mission within the Voyage 2050 planning cycle, 
covering the period from 2035 to 2050.
As a part of the ESA  Science Programme Voyage  2050, a new high--energy mission concept named 
\gq\ (Gamma Ray Astronomy International Laboratory for QUantum Exploration of Space--Time) has been presented in this paper.
 
The main scientific objectives that the mission aims to address are the following:
{\it i)} to localise Gamma--ray Burst's prompt emission with an accuracy of few arcseconds.
This capability is particularly relevant in light of the recent discovery that fast high energy transients are the electromagnetic counterparts of some gravitational wave events observed by the Advanced LIGO and Virgo network;
{\it ii)} to fully exploit timing capabilities down to micro--seconds or below at X/Gamma--ray energies, by means of an
adequate combination of temporal resolution and collecting area, thus allowing an effective investigation, for the first time, 
of the micro--second structure of Gamma--ray Bursts and other transient phenomena in the X/Gamma--ray energy window; 
 {\it iii)} to probe space--time structure down to the Planck scale 
by measuring the delays between photons of different energies in the prompt emission of Gamma--ray Bursts. 
More specifically, a significant class of theories of Quantum Gravity describing the
space--time structure down to the Planck scale predict a linear (w.r.t. photon energy) dispersion relation for light {\it in vacuo}. The predicted delays  
are tiny, being in the microsecond range, for photons of energies in the keV--MeV range, that travelled for a (few) billion years. In particular these effects scale almost linearly with the photon energy and the redshift of the GRB. 
This double linear dependence, in energy and redshift, is a unique signature of a Quantum Gravity effect, allowing for a robust experimental constraint within the proposed experiment.
  
\gq\ is a mission concept based on a constellation of nano/micro/small--satellites in low 
(or near) Earth orbits, hosting fast scintillators to probe the X/gamma--ray emission of bright
high--energy transients. The main features of this proposed experiment are: temporal resolution $\le 100$ nanoseconds, huge overall collecting area, $\sim 100$ square meters, very broad energy band coverage, $\sim 1$ keV--$10$ MeV.
\gq\ is conceived as an all--sky monitor for fast localisation of high signal--to--noise ratio transients in the broad 
keV--MeV band by robust triangulation techniques with accuracies at the micro--second level, and baselines of several thousand km. These features allow unprecedented localisation capabilities, in the keV--MeV band, of a few arcseconds or below, depending on the temporal structure of the transient event. Despite the huge collecting area, hundred(s) of square meters, and the consequent number of 
nano/micro/small--satellites utilised (from thousand(s) to ten(s)), all orbiting around the Earth in uniformly distributed orbits,
the technical capabilities and subsequent design of each base unit of the constellation are extremely simple and robust. 
This allows for mass--production of the base units of this experiment, namely a satellite equipped with a non--collimated (half--sky field of view) detector (effective area in the range hundred--thousand(s) square centimetres). The detector consists of segmented scintillator crystals coupled with Silicon Drift Detectors with broad energy band coverage (keV--MeV range) and excellent temporal resolution ($\le 100$ nanoseconds).
Although the field--of--view of the detectors is large ($\sim 2 \pi$ steradians), limited pointing capabilities are required. More specifically, instrument pointing at local zenith will not observe any Earth albedo from GRBs, which, otherwise, would greatly complicate the analysis. Nowadays, even with CubeSats, pointing accuracies of a few degrees are easily achievable. We forecast that mass production of this simple unit will allow a huge reduction of costs. Moreover, the large number of satellites involved in the \gq\ constellation make this experiment very robust against the failure of one or more of its units. 

\gq\ is a modular experiment in which, for each of the detected photons, only three measured parameters are essential, namely the accurate time--of--arrival of each photon (down to 100 nanoseconds, or below), the energy, with moderate resolution (few percent), and the detector position (within few tens of meters). 
This opens the compelling possibility of combining data from different kinds of detectors (aboard different kinds of satellites belonging, in principle, to different constellations) to achieve the scientific objectives of the \gq\ project, making \gq\ one of the few examples of modular space--based astronomy. Modular experiments have proven, in the past, to be very effective in opening up new possibilities for astronomical investigations. Just think of Very Large Baseline Interferometry, an astronomical interferometer in the Radio Band, involving more that thirty radio telescopes all over the world and Cluster II, a space mission of the European Space Agency, with NASA participation, composed of a constellation of four satellites, to study the Earth's magnetosphere, launched in 2000 and recently extended to the end of 2020. In the near future, a constellation of three satellites in formation is planned for the LISA mission, to study gravitational waves from space.
 Very recently, two extremely successful experiments, of paramount importance for fundamental physics, involve the combined use of several ground--based detectors. One is the LIGO/Virgo Collaboration (involving the two US--based LIGO and the European Virgo facilities) that gave us the first detection and localisation of gravitational waves. In one case, temporal triangulation techniques, conceptually similar to those proposed for the \gq\ constellation and described in this work, effectively constrained the position of the event in the sky, allowing for fast subsequent localisation, in the electromagnetic window, of a double Neutron Star merging event. The other is the Event Horizon Telescope (which utilizes 8 radio/micro--wave observatories spread all over the world) that obtained the first image of the event horizon around a black hole. We consider these compelling results as the proof that modular astronomy, which benefits from the combined use of distributed detectors (to increase the overall detecting area and allow for unprecedented spatial resolution, in the cases of the Event Horizon Telescope and the \gq\ project), is the new frontier of cutting--edge experimental astronomical science that is performed by exploiting the combination of a large number of detectors distributed all over the Earth's surface. The \gq\ project is a space--based version of this epochal revolution.
   
We performed accurate Monte--Carlo simulations of thousands of light curves of Gamma--ray Bursts, based on true data obtained from the scintillators of the Gamma Burst Monitor on board the {\it  Fermi} Satellite. We produced  Gamma--ray Burst light curves in consecutive energy bands in the interval 10 keV--50 MeV, for a range of effective areas. We then applied 
cross--correlation techniques to these light curves to determine the minimum accuracy with which potential temporal delays between these light curves are determined. As expected, this accuracy depends, in a complicated way, on the temporal variability scale of the Gamma--ray Burst considered, and scales roughly with the square root of the number of photons in the energy band considered. We determined that, for temporal variabilities in the millisecond range (which are expected in at least 30\% of the observed Gamma--ray Bursts), with an overall effective area of $\sim 100$ square meters, the statistical accuracy of these delays is always smaller (for redshifts $\ge 0.5$) than the delays expected in a dispersion law for the propagation of photons {\it in vacuo} that linearly depends on the ratio between the photon energy and the Planck energy.
 
This proves that the \gq\ constellation is able to achieve the ambitious objectives outlined above, within the budget of a European Space Agency M--class mission.


The biggest advantages of \gq\ with respect to standard All Sky Monitors for High Energy
Astrophysics are:

\begin{itemize}
\item
Modularity.
\item
Unprecedented temporal resolution.
\item
Limited cost and quick development.
\item
Huge effective area.
\end{itemize}

The first one allows us: a) to first fly a reduced version of \gq\ (say
4-12 units, the \gq\ pathfinder) to prove the concept (see also \S \, \ref{synergy} below); b) avoid
single (or even multiple) point failures: if one or several units are
lost the constellation and the experiment are not lost; c) initially test
the hardware with the first launches and then improve it, if needed,
with the following ones. 

The second allows \gq\ to open a new
window for studying micro--second variability in bright transients. 

To achieve the third characteristic  \gq\
will exploit commercial off--the--shelf hardware as well as the trend in
reducing the cost of both manufacturing and launching of micro/nano--satellites
over the next years. \gq\ would naturally fit into a scheme where
production of identical units would follow the development and
testing of a first {\it test unit}. The development of engineering
and qualification models, and all tests at the level of critical
components, will be performed only for the {\it test unit}. For the
other units only flight models will be built, and these units
will be tested only at the system level. All this will bring costs
down and speed up the construction of the full mission.

Finally, in view of the limited costs and quick development, it is possible to build an all--sky monitor of 
unprecedented area ($\sim 100 \, {\rm m}^2$). 
The consequent sensitivity to extremely weak transients 
is mandatory to fully exploit the exciting possibilities offered by the birth of Multi-Messenger Astronomy. 
Starting in 2025 the improved or next generation of gravitational wave detectors LIGO--Virgo, KAGRA, and the Einstein Telescope will provide
detectability of NS--NS mergers events like GW170817 within a few hundred Mpc. This corresponds to faint electromagnetic counterparts that
require high--sensitivity all--sky monitors to be effectively detected and studied.
Moreover, the extraordinary number of photons detected with astonishing temporal accuracy from each GRB, will allow us, at least for the brightest events, 
to perform the first dedicated experiment in Quantum Gravity to test, with meaningful accuracy, a first order dispersion relation for light {\it in vacuo}. In this respect 
\gq\ will be the first experiment potentially able to reveal a Space--Time granularity at the minuscule Planck length scale.

\section{Synergy with other on going projects}
\label{synergy}

Some of the authors of this paper are developing the High Energy Rapid
Modular Ensemble of Satellites, HERMES, pathfinder experiment.  HERMES
pathfinder consists of six nano--satellites of the 3U class each equipped with a payload consisting of GAGG scintillators coupled with SDDs
with a collecting area of about 55 cm$^2$ per payload. The main goals of
the HERMES pathfinder are to prove that GRB prompt events can be
efficiently and routinely observed with detectors hosted by
nano--satellites, and to test GRB localisation techniques based on triangulation using the delays of photon arrival times 
on different detectors located in low Earth orbit.  
The HERMES pathfinder experiment will test fast timing
techniques that are at the core of the GrailQuest project. The design
performance of the HERMES pathfinder detectors guarantee a temporal resolution of 300
nanoseconds, 5--10 times better than most current and past GRB
experiments.  HERMES pathfinder is funded by the Italian Space Agency
and by the European Community through the HERMES-SP H2020 SPACE
grant. More information on the HERMES pathfinder can be found at
www.hermes-sp.eu and hermes.dsf.unica.it.



\begin{acknowledgements}
Some of the authors wish to thank ASI and INAF (agreements ASI-UNI-Ca 2016-13-U.O and ASI- INAF 2018-10-hh.0), MIUR, Italy (HERMES-TP project) and EU (HERMES-SP Horizon 2020 Research and Innovation Project under grant agreement 821896) for financial support within the HERMES project.
\end{acknowledgements}

%
%

\bibliographystyle{spbasic}      
\bibliography{biblio}   

\begin{thebibliography}{82}
\providecommand{\natexlab}[1]{#1}
\providecommand{\url}[1]{{#1}}
\providecommand{\urlprefix}{URL }
\expandafter\ifx\csname urlstyle\endcsname\relax
  \providecommand{\doi}[1]{DOI~\discretionary{}{}{}#1}\else
  \providecommand{\doi}{DOI~\discretionary{}{}{}\begingroup
  \urlstyle{rm}\Url}\fi
\providecommand{\eprint}[2][]{\url{#2}}

\bibitem[{{Abbott} et~al.(2016){Abbott}, {Abbott}, {Abbott}, {Abernathy},
  {Acernese}, and {Ackley}}]{GW150914}
{Abbott} BP, {Abbott} R, {Abbott} TD, {Abernathy} MR, {Acernese} F, {Ackley} ea
  (2016) {Observation of Gravitational Waves from a Binary Black Hole Merger}.
  Physical Review Letters 116(6):061102, \doi{10.1103/PhysRevLett.116.061102},
  \eprint{1602.03837}

\bibitem[{{Abbott} et~al.(2017{\natexlab{a}}){Abbott}, {Abbott}, {Abbott},
  {Acernese}, {Ackley}, and {Adams}}]{GW170104}
{Abbott} BP, {Abbott} R, {Abbott} TD, {Acernese} F, {Ackley} K, {Adams} Cea
  (2017{\natexlab{a}}) {GW170104: Observation of a 50-Solar-Mass Binary Black
  Hole Coalescence at Redshift 0.2}. Physical Review Letters 118(22):221101,
  \doi{10.1103/PhysRevLett.118.221101}, \eprint{1706.01812}

\bibitem[{{Abbott} et~al.(2017{\natexlab{b}}){Abbott}, {Abbott}, {Abbott},
  {Acernese}, {Ackley}, and {Adams}}]{GW170814}
{Abbott} BP, {Abbott} R, {Abbott} TD, {Acernese} F, {Ackley} K, {Adams} Cea
  (2017{\natexlab{b}}) {GW170814: A Three-Detector Observation of Gravitational
  Waves from a Binary Black Hole Coalescence}. Physical Review Letters
  119(14):141101, \doi{10.1103/PhysRevLett.119.141101}, \eprint{1709.09660}

\bibitem[{{Abbott} et~al.(2017{\natexlab{c}}){Abbott}, {Abbott}, {Abbott},
  {Acernese}, {Ackley}, and {Adams}}]{GW170817}
{Abbott} BP, {Abbott} R, {Abbott} TD, {Acernese} F, {Ackley} K, {Adams} Cea
  (2017{\natexlab{c}}) {GW170817: Observation of Gravitational Waves from a
  Binary Neutron Star Inspiral}. Physical Review Letters 119(16):161101,
  \doi{10.1103/PhysRevLett.119.161101}, \eprint{1710.05832}

\bibitem[{{Abbott} et~al.(2017{\natexlab{d}}){Abbott}, {Abbott}, {Abbott},
  {Acernese}, {Ackley}, and {Adams}}]{GW170817_GW}
{Abbott} BP, {Abbott} R, {Abbott} TD, {Acernese} F, {Ackley} K, {Adams} ea C
  (2017{\natexlab{d}}) {Gravitational Waves and Gamma-Rays from a Binary
  Neutron Star Merger: GW170817 and GRB 170817A}. Astrophysical Journal,
  Letters 848(2):L13, \doi{10.3847/2041-8213/aa920c}, \eprint{1710.05834}

\bibitem[{Abbott et~al.(2017)Abbott, Abbott, Abbott, Acernese, Ackley, and
  et~al.}]{Abbot_mu_2017}
Abbott BP, Abbott R, Abbott TD, Acernese F, Ackley K, et~al CA (2017)
  Multi-messenger observations of a binary neutron star merger. The
  Astrophysical Journal 848(2):L12, \doi{10.3847/2041-8213/aa91c9}

\bibitem[{{Abdo} et~al.(2009){Abdo}, {Ackermann}, {Ajello}, and
  {Asano}}]{Abdo09b}
{Abdo} AA, {Ackermann} M, {Ajello} M, {Asano} ea (2009) {A limit on the
  variation of the speed of light arising from quantum gravity effects}. Nature
  462(7271):331--334, \doi{10.1038/nature08574}, \eprint{0908.1832}

\bibitem[{{Amelino-Camelia}(2000)}]{Amelino00}
{Amelino-Camelia} G (2000) {Are We at the Dawn of Quantum-Gravity
  Phenomenology?}, vol 541, p~1

\bibitem[{{Amelino-Camelia} and {Smolin}(2009)}]{Amelino09}
{Amelino-Camelia} G, {Smolin} L (2009) {Prospects for constraining quantum
  gravity dispersion with near term observations}. Physical Review D
  80(8):084017, \doi{10.1103/PhysRevD.80.084017}, \eprint{0906.3731}

\bibitem[{{Amelino-Camelia} et~al.(1998){Amelino-Camelia}, {Ellis},
  {Mavromatos}, {Nanopoulos}, and {Sarkar}}]{Camelia98}
{Amelino-Camelia} G, {Ellis} J, {Mavromatos} NE, {Nanopoulos} DV, {Sarkar} S
  (1998) {Tests of quantum gravity from observations of
  {\ensuremath{\gamma}}-ray bursts}. Nature 393(6687):763--765,
  \doi{10.1038/31647}, \eprint{astro-ph/9712103}

\bibitem[{{Amelino-Camelia} et~al.(2017){Amelino-Camelia}, {D'Amico}, {Rosati},
  and {Loret}}]{Amelino17a}
{Amelino-Camelia} G, {D'Amico} G, {Rosati} G, {Loret} N (2017) {In vacuo
  dispersion features for gamma-ray-burst neutrinos and photons}. Nature
  Astronomy 1:0139, \doi{10.1038/s41550-017-0139}, \eprint{1612.02765}

\bibitem[{{Ashtekar} and {Lewandowski}(1997)}]{Astekhar97}
{Ashtekar} A, {Lewandowski} J (1997) {Quantum theory of geometry: I. Area
  operators}. Classical and Quantum Gravity 14(1A):A55--A81,
  \doi{10.1088/0264-9381/14/1A/006}, \eprint{gr-qc/9602046}

\bibitem[{{Band} et~al.(1993){Band}, {Matteson}, {Ford}, {Schaefer}, {Palmer},
  {Teegarden}, {Cline}, {Briggs}, {Paciesas}, {Pendleton}, {Fishman},
  {Kouveliotou}, {Meegan}, {Wilson}, and {Lestrade}}]{Band93}
{Band} D, {Matteson} J, {Ford} L, {Schaefer} B, {Palmer} D, {Teegarden} B,
  {Cline} T, {Briggs} M, {Paciesas} W, {Pendleton} G, {Fishman} G,
  {Kouveliotou} C, {Meegan} C, {Wilson} R, {Lestrade} P (1993) {BATSE
  Observations of Gamma-Ray Burst Spectra. I. Spectral Diversity}.
  Astrophysical Journal 413:281, \doi{10.1086/172995}

\bibitem[{{Barrau} et~al.(2014){Barrau}, {Rovelli}, and {Vidotto}}]{Barrau14}
{Barrau} A, {Rovelli} C, {Vidotto} F (2014) {Fast radio bursts and white hole
  signals}. Physical Review D 90(12):127503, \doi{10.1103/PhysRevD.90.127503},
  \eprint{1409.4031}

\bibitem[{{Barthelmy}(2004)}]{Barthelmy2004}
{Barthelmy} SD (2004) {Burst Alert Telescope (BAT) on the Swift MIDEX mission}.
  In: {Flanagan} KA, {Siegmund} OHW (eds) X-Ray and Gamma-Ray Instrumentation
  for Astronomy XIII, Society of Photo-Optical Instrumentation Engineers (SPIE)
  Conference Series, vol 5165, pp 175--189, \doi{10.1117/12.506779}

\bibitem[{{Bhat} et~al.(2012){Bhat}, {Briggs}, {Connaughton}, {Kouveliotou},
  {van der Horst}, {Paciesas}, {Meegan}, {Bissaldi}, {Burgess}, {Chaplin},
  {Diehl}, {Fishman}, {Fitzpatrick}, {Foley}, {Gibby}, {Giles}, {Goldstein},
  {Greiner}, {Gruber}, {Guiriec}, {von Kienlin}, {Kippen}, {McBreen}, {Preece},
  {Rau}, {Tierney}, and {Wilson-Hodge}}]{Bhat12}
{Bhat} PN, {Briggs} MS, {Connaughton} V, {Kouveliotou} C, {van der Horst} AJ,
  {Paciesas} W, {Meegan} CA, {Bissaldi} E, {Burgess} M, {Chaplin} V, {Diehl} R,
  {Fishman} G, {Fitzpatrick} G, {Foley} S, {Gibby} M, {Giles} MM, {Goldstein}
  A, {Greiner} J, {Gruber} D, {Guiriec} S, {von Kienlin} A, {Kippen} M,
  {McBreen} S, {Preece} R, {Rau} A, {Tierney} D, {Wilson-Hodge} C (2012)
  {Temporal Deconvolution Study of Long and Short Gamma-Ray Burst Light
  Curves}. Astrophysical Journal 744(2):141, \doi{10.1088/0004-637X/744/2/141},
  \eprint{1109.4064}

\bibitem[{{Bloom}(2011)}]{Bloom11}
{Bloom} DE (2011) {7 Billion and Counting}. Science 333(6042):562,
  \doi{10.1126/science.1209290}

\bibitem[{{Bloom} et~al.(2009){Bloom}, {Perley}, {Li}, {Butler}, {Miller},
  {Kocevski}, {Kann}, {Foley}, {Chen}, {Filippenko}, {Starr}, {Macomber},
  {Prochaska}, {Chornock}, {Poznanski}, {Klose}, {Skrutskie}, {Lopez}, {Hall},
  {Glazebrook}, and {Blake}}]{Bloom09}
{Bloom} JS, {Perley} DA, {Li} W, {Butler} NR, {Miller} AA, {Kocevski} D, {Kann}
  DA, {Foley} RJ, {Chen} HW, {Filippenko} AV, {Starr} DL, {Macomber} B,
  {Prochaska} JX, {Chornock} R, {Poznanski} D, {Klose} S, {Skrutskie} MF,
  {Lopez} S, {Hall} P, {Glazebrook} K, {Blake} CH (2009) {Observations of the
  Naked-Eye GRB 080319B: Implications of Nature's Brightest Explosion}.
  Astrophysical Journal 691(1):723--737, \doi{10.1088/0004-637X/691/1/723},
  \eprint{0803.3215}

\bibitem[{Bridgman(1927)}]{Bridgman27}
Bridgman P (1927) The Logic of Modern Physics. Macmillan paperbacks, Macmillan,
  \urlprefix\url{https://books.google.it/books?id=zGUGAQAAIAAJ}

\bibitem[{{Burderi} et~al.(2016){Burderi}, {Di Salvo}, and {Iaria}}]{Burderi16}
{Burderi} L, {Di Salvo} T, {Iaria} R (2016) {Quantum clock: A critical
  discussion on spacetime}. Physical Review D 93(6):064017,
  \doi{10.1103/PhysRevD.93.064017}, \eprint{1603.03723}

\bibitem[{{Campana} and {Di Salvo}(2018)}]{DiSalvo_2018}
{Campana} S, {Di Salvo} T (2018) {Accreting Pulsars: Mixing-up Accretion Phases
  in Transitional Systems}, Astrophysics and Space Science Library, vol 457, p
  149

\bibitem[{{Capozziello} and {de Laurentis}(2011)}]{CapozzielloDeLaurentis2011}
{Capozziello} S, {de Laurentis} M (2011) {Extended Theories of Gravity}.
  Physics Reports 509(4):167--321, \doi{10.1016/j.physrep.2011.09.003},
  \eprint{1108.6266}

\bibitem[{{Capozziello} and {Lambiase}(2015)}]{Capozziello15}
{Capozziello} S, {Lambiase} G (2015) {The emission of Gamma Ray Bursts as a
  test-bed for modified gravity}. Physics Letters B 750:344--347,
  \doi{10.1016/j.physletb.2015.09.048}, \eprint{1504.03900}

\bibitem[{{Chan} et~al.(2018){Chan}, {Messenger}, {Heng}, and
  {Hendry}}]{Chan18}
{Chan} ML, {Messenger} C, {Heng} IS, {Hendry} M (2018) {Binary neutron star
  mergers and third generation detectors: Localization and early warning}.
  Physical Review D 97(12):123014, \doi{10.1103/PhysRevD.97.123014},
  \eprint{1803.09680}

\bibitem[{{Costa} et~al.(1997){Costa}, {Frontera}, {Heise}, {Feroci}, {in't
  Zand}, {Fiore}, {Cinti}, {Dal Fiume}, {Nicastro}, {Orlandini}, {Palazzi},
  {Rapisarda\#}, {Zavattini}, {Jager}, {Parmar}, {Owens}, {Molendi},
  {Cusumano}, {Maccarone}, {Giarrusso}, {Coletta}, {Antonelli}, {Giommi},
  {Muller}, {Piro}, and {Butler}}]{Costa97}
{Costa} E, {Frontera} F, {Heise} J, {Feroci} M, {in't Zand} J, {Fiore} F,
  {Cinti} MN, {Dal Fiume} D, {Nicastro} L, {Orlandini} M, {Palazzi} E,
  {Rapisarda\#} M, {Zavattini} G, {Jager} R, {Parmar} A, {Owens} A, {Molendi}
  S, {Cusumano} G, {Maccarone} MC, {Giarrusso} S, {Coletta} A, {Antonelli} LA,
  {Giommi} P, {Muller} JM, {Piro} L, {Butler} RC (1997) {Discovery of an X-ray
  afterglow associated with the {\ensuremath{\gamma}}-ray burst of 28 February
  1997}. Nature 387(6635):783--785, \doi{10.1038/42885},
  \eprint{astro-ph/9706065}

\bibitem[{{Cucchiara} et~al.(2011){Cucchiara}, {Levan}, {Fox}, {Tanvir},
  {Ukwatta}, {Berger}, {Kr{\"u}hler}, {K{\"u}pc{\"u} Yolda{\textcommabelow s}},
  {Wu}, {Toma}, {Greiner}, {Olivares}, {Rowlinson}, {Amati}, {Sakamoto},
  {Roth}, {Stephens}, {Fritz}, {Fynbo}, {Hjorth}, {Malesani}, {Jakobsson},
  {Wiersema}, {O'Brien}, {Soderberg}, {Foley}, {Fruchter}, {Rhoads},
  {Rutledge}, {Schmidt}, {Dopita}, {Podsiadlowski}, {Willingale}, {Wolf},
  {Kulkarni}, and {D'Avanzo}}]{Cucchiara_2011}
{Cucchiara} A, {Levan} AJ, {Fox} DB, {Tanvir} NR, {Ukwatta} TN, {Berger} E,
  {Kr{\"u}hler} T, {K{\"u}pc{\"u} Yolda{\textcommabelow s}} A, {Wu} XF, {Toma}
  K, {Greiner} J, {Olivares} FE, {Rowlinson} A, {Amati} L, {Sakamoto} T, {Roth}
  K, {Stephens} A, {Fritz} A, {Fynbo} JPU, {Hjorth} J, {Malesani} D,
  {Jakobsson} P, {Wiersema} K, {O'Brien} PT, {Soderberg} AM, {Foley} RJ,
  {Fruchter} AS, {Rhoads} J, {Rutledge} RE, {Schmidt} BP, {Dopita} MA,
  {Podsiadlowski} P, {Willingale} R, {Wolf} C, {Kulkarni} SR, {D'Avanzo} P
  (2011) {A Photometric Redshift of z \raisebox{-0.5ex}\textasciitilde 9.4 for
  GRB 090429B}. Astrophysical Journal 736(1):7,
  \doi{10.1088/0004-637X/736/1/7}, \eprint{1105.4915}

\bibitem[{{Doplicher} et~al.(1995){Doplicher}, {Fredenhagen}, and
  {Roberts}}]{Doplicher95}
{Doplicher} S, {Fredenhagen} K, {Roberts} JE (1995) {The quantum structure of
  spacetime at the Planck scale and quantum fields}. Communications in
  Mathematical Physics 172(1):187--220, \doi{10.1007/BF02104515},
  \eprint{hep-th/0303037}

\bibitem[{{Einstein}(1905{\natexlab{a}})}]{EinsteinI}
{Einstein} A (1905{\natexlab{a}}) {{\"U}ber die von der molekularkinetischen
  Theorie der W{\"a}rme geforderte Bewegung von in ruhenden Fl{\"u}ssigkeiten
  suspendierten Teilchen}. Annalen der Physik 322(8):549--560,
  \doi{10.1002/andp.19053220806}

\bibitem[{{Einstein}(1905{\natexlab{b}})}]{EinsteinII}
{Einstein} A (1905{\natexlab{b}}) {{\"U}ber einen die Erzeugung und Verwandlung
  des Lichtes betreffenden heuristischen Gesichtspunkt}. Annalen der Physik
  322(6):132--148, \doi{10.1002/andp.19053220607}

\bibitem[{{Ellis} et~al.(2019){Ellis}, {Konoplich}, {Mavromatos}, {Nguyen},
  {Sakharov}, and {Sarkisyan-Grinbaum}}]{Ellis19}
{Ellis} J, {Konoplich} R, {Mavromatos} NE, {Nguyen} L, {Sakharov} AS,
  {Sarkisyan-Grinbaum} EK (2019) {Robust constraint on Lorentz violation using
  Fermi-LAT gamma-ray burst data}. Physical Review D 99(8):083009,
  \doi{10.1103/PhysRevD.99.083009}, \eprint{1807.00189}

\bibitem[{{Fiore} et~al.(2000){Fiore}, {Nicastro}, {Savaglio}, {Stella}, and
  {Vietri}}]{Fiore00}
{Fiore} F, {Nicastro} F, {Savaglio} S, {Stella} L, {Vietri} M (2000) {Probing
  the Warm Intergalactic Medium through Absorption against Gamma-Ray Burst
  X-Ray Afterglows}. Astrophysical Journal, Letters 544(1):L7--L10,
  \doi{10.1086/317286}, \eprint{astro-ph/0009292}

\bibitem[{{Garay}(1995)}]{Garay95}
{Garay} LJ (1995) {Quantum Gravity and Minimum Length}. International Journal
  of Modern Physics A 10(2):145--165, \doi{10.1142/S0217751X95000085},
  \eprint{gr-qc/9403008}

\bibitem[{{Hentschel} et~al.(2001){Hentschel}, {Kienberger}, {Spielmann},
  {Reider}, N., T., {Heinzmann}, and {Krausz}}]{Hentschel01}
{Hentschel} M, {Kienberger} R, {Spielmann} C, {Reider} GAM, N B, T P {Corkum},
  {Heinzmann} M Uand~{Drescher}, {Krausz} F (2001) {Attosecond metrology}.
  Nature 414:509--513, \doi{10.1038/35107000}

\bibitem[{{Holoien} et~al.(2016){Holoien}, {Kochanek}, {Prieto}, {Stanek},
  {Dong}, {Shappee}, {Grupe}, {Brown}, {Basu}, {Beacom}, {Bersier},
  {Brimacombe}, {Danilet}, {Falco}, {Guo}, {Jose}, {Herczeg}, {Long},
  {Pojmanski}, {Simonian}, {Szczygie{\l}}, {Thompson}, {Thorstensen}, {Wagner},
  and {Wo{\'z}niak}}]{Holoien16}
{Holoien} TWS, {Kochanek} CS, {Prieto} JL, {Stanek} KZ, {Dong} S, {Shappee} BJ,
  {Grupe} D, {Brown} JS, {Basu} U, {Beacom} JF, {Bersier} D, {Brimacombe} J,
  {Danilet} AB, {Falco} E, {Guo} Z, {Jose} J, {Herczeg} GJ, {Long} F,
  {Pojmanski} G, {Simonian} GV, {Szczygie{\l}} DM, {Thompson} TA, {Thorstensen}
  JR, {Wagner} RM, {Wo{\'z}niak} PR (2016) {Six months of multiwavelength
  follow-up of the tidal disruption candidate ASASSN-14li and implied TDE rates
  from ASAS-SN}. Monthly Notices of the RAS 455(3):2918--2935,
  \doi{10.1093/mnras/stv2486}, \eprint{1507.01598}

\bibitem[{{Hossenfelder}(2013)}]{Hossenfelder12}
{Hossenfelder} S (2013) {Minimal Length Scale Scenarios for Quantum Gravity}.
  Living Reviews in Relativity 16(1):2, \doi{10.12942/lrr-2013-2},
  \eprint{1203.6191}

\bibitem[{{Jacob} and {Piran}(2008)}]{Jacob08}
{Jacob} U, {Piran} T (2008) {Lorentz-violation-induced arrival delays of
  cosmological particles}. Journal of Cosmology and Astroparticle Physics
  2008(1):031, \doi{10.1088/1475-7516/2008/01/031}, \eprint{0712.2170}

\bibitem[{{Kobayashi} et~al.(1997){Kobayashi}, {Piran}, and
  {Sari}}]{Kobayashi97}
{Kobayashi} S, {Piran} T, {Sari} R (1997) {Can Internal Shocks Produce the
  Variability in Gamma-Ray Bursts?} Astrophysical Journal 490:92,
  \doi{10.1086/512791}, \eprint{astro-ph/9705013}

\bibitem[{{Kobayashi} et~al.(2004){Kobayashi}, {Laguna}, {Phinney}, and
  {M{\'e}sz{\'a}ros}}]{Kobayashi04}
{Kobayashi} S, {Laguna} P, {Phinney} ES, {M{\'e}sz{\'a}ros} P (2004)
  {Gravitational Waves and X-Ray Signals from Stellar Disruption by a Massive
  Black Hole}. Astrophysical Journal 615(2):855--865, \doi{10.1086/424684},
  \eprint{astro-ph/0404173}

\bibitem[{{Liberati}(2013)}]{Liberati13}
{Liberati} S (2013) {Tests of Lorentz invariance: a 2013 update}. Classical and
  Quantum Gravity 30(13):133001, \doi{10.1088/0264-9381/30/13/133001},
  \eprint{1304.5795}

\bibitem[{{Lorentz} and {Brun}(2017)}]{Lorentz17}
{Lorentz} M, {Brun} P (2017) {Limits on Lorentz invariance violation at the
  Planck energy scale from H.E.S.S. spectral analysis of the blazar Mrk 501}.
  In: European Physical Journal Web of Conferences, European Physical Journal
  Web of Conferences, vol 136, p 03018, \doi{10.1051/epjconf/201713603018},
  \eprint{1606.08600}

\bibitem[{{Lorimer}(2018)}]{Lorimer_2018}
{Lorimer} DR (2018) {A decade of fast radio bursts}. Nature Astronomy
  2:860--864, \doi{10.1038/s41550-018-0607-9}, \eprint{1811.00195}

\bibitem[{{MacLachlan} et~al.(2013){MacLachlan}, {Shenoy}, {Sonbas}, {Dhuga},
  {Cobb}, {Ukwatta}, {Morris}, {Eskandarian}, {Maximon}, and
  {Parke}}]{MacLachlan13}
{MacLachlan} GA, {Shenoy} A, {Sonbas} E, {Dhuga} KS, {Cobb} BE, {Ukwatta} TN,
  {Morris} DC, {Eskandarian} A, {Maximon} LC, {Parke} WC (2013) {Minimum
  variability time-scales of long and short GRBs}. Monthly Notices of the RAS
  432(2):857--865, \doi{10.1093/mnras/stt241}, \eprint{1201.4431}

\bibitem[{{Mattingly}(2005)}]{Mattingly05}
{Mattingly} D (2005) {Modern Tests of Lorentz Invariance}. Living Reviews in
  Relativity 8(1):5, \doi{10.12942/lrr-2005-5}, \eprint{gr-qc/0502097}

\bibitem[{{McConnell} et~al.(1996){McConnell}, {Forrest}, {Vestrand}, and
  {Finger}}]{McConnell_1996}
{McConnell} M, {Forrest} D, {Vestrand} WT, {Finger} M (1996) {Using BATSE to
  measure gamma-ray burst polarization}. In: {Kouveliotou} C, {Briggs} MF,
  {Fishman} GJ (eds) American Institute of Physics Conference Series, American
  Institute of Physics Conference Series, vol 384, pp 851--855,
  \doi{10.1063/1.51605}

\bibitem[{Mead(1964)}]{Mead64}
Mead CA (1964) Possible connection between gravitation and fundamental length.
  Physical Review (US) Superseded in part by Phys Rev A, Phys Rev B: Solid
  State, Phys Rev C, and Phys Rev D 135, \doi{10.1103/PhysRev.135.B849}

\bibitem[{{Meegan} et~al.(2009){Meegan}, {Lichti}, {Bhat}, {Bissaldi},
  {Briggs}, {Connaughton}, {Diehl}, {Fishman}, {Greiner}, {Hoover}, {van der
  Horst}, {von Kienlin}, {Kippen}, {Kouveliotou}, {McBreen}, {Paciesas},
  {Preece}, {Steinle}, {Wallace}, {Wilson}, and {Wilson-Hodge}}]{Meegan09}
{Meegan} C, {Lichti} G, {Bhat} PN, {Bissaldi} E, {Briggs} MS, {Connaughton} V,
  {Diehl} R, {Fishman} G, {Greiner} J, {Hoover} AS, {van der Horst} AJ, {von
  Kienlin} A, {Kippen} RM, {Kouveliotou} C, {McBreen} S, {Paciesas} WS,
  {Preece} R, {Steinle} H, {Wallace} MS, {Wilson} RB, {Wilson-Hodge} C (2009)
  {The Fermi Gamma-ray Burst Monitor}. Astrophysical Journal 702(1):791--804,
  \doi{10.1088/0004-637X/702/1/791}, \eprint{0908.0450}

\bibitem[{{Metzger} et~al.(1997){Metzger}, {Djorgovski}, {Kulkarni}, {Steidel},
  {Adelberger}, {Frail}, {Costa}, and {Frontera}}]{Metzger97}
{Metzger} MR, {Djorgovski} SG, {Kulkarni} SR, {Steidel} CC, {Adelberger} KL,
  {Frail} DA, {Costa} E, {Frontera} F (1997) {Spectral constraints on the
  redshift of the optical counterpart to the {\ensuremath{\gamma}}-ray burst of
  8 May 1997}. Nature 387(6636):878--880, \doi{10.1038/43132}

\bibitem[{{Michelson} and {Morley}(1887)}]{Michelson87}
{Michelson} AA, {Morley} EW (1887) {On the relative motion of the Earth and the
  luminiferous ether}. American Journal of Science 34(203):333--345,
  \doi{10.2475/ajs.s3-34.203.333}

\bibitem[{{Morsony} et~al.(2010){Morsony}, {Lazzati}, and
  {Begelman}}]{Morsony10}
{Morsony} BJ, {Lazzati} D, {Begelman} MC (2010) {The Origin and Propagation of
  Variability in the Outflows of Long-duration Gamma-ray Bursts}. Astrophysical
  Journal 723(1):267--276, \doi{10.1088/0004-637X/723/1/267},
  \eprint{1002.0361}

\bibitem[{{Nakar} and {Piran}(2002)}]{Nakar02}
{Nakar} E, {Piran} T (2002) {Gamma-Ray Burst Light Curves-Another Clue on the
  Inner Engine}. Astrophysical Journal, Letters 572(2):L139--L142,
  \doi{10.1086/341748}, \eprint{astro-ph/0202404}

\bibitem[{{Pasham} et~al.(2019){Pasham}, {Remillard}, {Fragile}, {Franchini},
  {Stone}, {Lodato}, {Homan}, {Chakrabarty}, {Baganoff}, {Steiner}, {Coughlin},
  and {Pasham}}]{pasham19}
{Pasham} DR, {Remillard} RA, {Fragile} PC, {Franchini} A, {Stone} NC, {Lodato}
  G, {Homan} J, {Chakrabarty} D, {Baganoff} FK, {Steiner} JF, {Coughlin} ER,
  {Pasham} NR (2019) {A loud quasi-periodic oscillation after a star is
  disrupted by a massive black hole}. Science 363(6426):531--534,
  \doi{10.1126/science.aar7480}, \eprint{1810.10713}

\bibitem[{{Penrose}(1969)}]{Penrose69}
{Penrose} R (1969) {Gravitational Collapse: the Role of General Relativity}.
  Nuovo Cimento Rivista Serie 1:252

\bibitem[{{Phinney}(2009)}]{Phinney09}
{Phinney} ES (2009) {Finding and Using Electromagnetic Counterparts of
  Gravitational Wave Sources}. In: astro2010: The Astronomy and Astrophysics
  Decadal Survey, vol 2010, p 235, \eprint{0903.0098}

\bibitem[{{Planck}(1900)}]{Planck99}
{Planck} M (1900) {Ueber irreversible Strahlungsvorg{\"a}nge}. Annalen der
  Physik 306(1):69--122, \doi{10.1002/andp.19003060105}

\bibitem[{Plato(127b-e)}]{Plato}
Plato (127b-e) Parmenides

\bibitem[{{Ramirez-Ruiz} and {Fenimore}(2000)}]{Ramirez00}
{Ramirez-Ruiz} E, {Fenimore} EE (2000) {Pulse Width Evolution in Gamma-Ray
  Bursts: Evidence for Internal Shocks}. Astrophysical Journal 539(2):712--717,
  \doi{10.1086/309260}, \eprint{astro-ph/9910273}

\bibitem[{{Rees}(1988)}]{Rees88}
{Rees} MJ (1988) {Tidal disruption of stars by black holes of
  {}10$^{6}$-{}10$^{8}$ solar masses in nearby galaxies}. Nature
  333(6173):523--528, \doi{10.1038/333523a0}

\bibitem[{{Rosati} et~al.(2015){Rosati}, {Amelino-Camelia}, {Marcian{\`o}}, and
  {Matassa}}]{Rosati15}
{Rosati} G, {Amelino-Camelia} G, {Marcian{\`o}} A, {Matassa} M (2015)
  {Planck-scale-modified dispersion relations in FRW spacetime}. Physical
  Review D 92(12):124042, \doi{10.1103/PhysRevD.92.124042}, \eprint{1507.02056}

\bibitem[{{Rovelli}(1993)}]{Rovelli93}
{Rovelli} C (1993) {A generally covariant quantum field theory and a prediction
  on quantum measurements of geometry}. Nuclear Physics B 405(2):797--815,
  \doi{10.1016/0550-3213(93)90567-9}

\bibitem[{{Rovelli}(1998)}]{Rovelli98}
{Rovelli} C (1998) {Loop Quantum Gravity}. Living Reviews in Relativity 1(1):1,
  \doi{10.12942/lrr-1998-1}, \eprint{gr-qc/9710008}

\bibitem[{{Rovelli} and {Smolin}(1988)}]{Rovelli88a}
{Rovelli} C, {Smolin} L (1988) {Knot theory and quantum gravity}. Physical
  Review Letters 61:1155--1158, \doi{10.1103/PhysRevLett.61.1155}

\bibitem[{{Rovelli} and {Smolin}(1990)}]{Rovelli88b}
{Rovelli} C, {Smolin} L (1990) {Loop space representation of quantum general
  relativity}. Nuclear Physics B 331(1):80--152,
  \doi{10.1016/0550-3213(90)90019-A}

\bibitem[{{Rovelli} and {Smolin}(1995)}]{Rovelli95}
{Rovelli} C, {Smolin} L (1995) {Discreteness of area and volume in quantum
  gravity}. Nuclear Physics B 442(3):593--619,
  \doi{10.1016/0550-3213(95)00150-Q}, \eprint{gr-qc/9411005}

\bibitem[{{Rovelli} and {Speziale}(2003)}]{Rovelli03}
{Rovelli} C, {Speziale} S (2003) {Reconcile Planck-scale discreteness and the
  Lorentz-Fitzgerald contraction}. Physical Review D 67(6):064019,
  \doi{10.1103/PhysRevD.67.064019}, \eprint{gr-qc/0205108}

\bibitem[{Russell(1903)}]{Russell}
Russell B (1903) The Principles of Mathematics. Cambridge University Press,
  Cambridge, \urlprefix\url{https://books.google.it/books?id=kj0a\_aV2mxIC}

\bibitem[{{Schneider} et~al.(2002){Schneider}, {Guetta}, and
  {Ferrara}}]{Schneider02}
{Schneider} R, {Guetta} D, {Ferrara} A (2002) {Gamma-ray bursts from the first
  stars: neutrino signals}. Monthly Notices of the RAS 334(1):173--181,
  \doi{10.1046/j.1365-8711.2002.05511.x}, \eprint{astro-ph/0201342}

\bibitem[{Smolin(2007)}]{smolin2007}
Smolin L (2007) The Trouble with Physics: The Rise of String Theory, the Fall
  of a Science, and What Comes Next. Houghton Mifflin Harcourt,
  \urlprefix\url{https://books.google.it/books?id=d6MIUlxY-qwC}

\bibitem[{{Spada} et~al.(2000){Spada}, {Panaitescu}, and
  {M{\'e}sz{\'a}ros}}]{Spada00}
{Spada} M, {Panaitescu} A, {M{\'e}sz{\'a}ros} P (2000) {Analysis of Temporal
  Features of Gamma-Ray Bursts in the Internal Shock Model}. Astrophysical
  Journal 537(2):824--832, \doi{10.1086/309048}, \eprint{astro-ph/9908097}

\bibitem[{Susskind(2008)}]{Susskind}
Susskind L (2008) The Black Hole War. Little, Brown and Company

\bibitem[{{Troja} et~al.(2017){Troja}, {Piro}, {van Eerten}, {Wollaeger}, {Im},
  {Fox}, {Butler}, {Cenko}, {Sakamoto}, {Fryer}, {Ricci}, {Lien}, {Ryan},
  {Korobkin}, {Lee}, {Burgess}, {Lee}, {Watson}, {Choi}, {Covino}, {D'Avanzo},
  {Fontes}, {Gonz{\'a}lez}, {Khandrika}, {Kim}, {Kim}, {Lee}, {Lee}, {Kutyrev},
  {Lim}, {S{\'a}nchez-Ram{\'\i}rez}, {Veilleux}, {Wieringa}, and
  {Yoon}}]{Troja17}
{Troja} E, {Piro} L, {van Eerten} H, {Wollaeger} RT, {Im} M, {Fox} OD, {Butler}
  NR, {Cenko} SB, {Sakamoto} T, {Fryer} CL, {Ricci} R, {Lien} A, {Ryan} RE,
  {Korobkin} O, {Lee} SK, {Burgess} JM, {Lee} WH, {Watson} AM, {Choi} C,
  {Covino} S, {D'Avanzo} P, {Fontes} CJ, {Gonz{\'a}lez} JB, {Khandrika} HG,
  {Kim} J, {Kim} SL, {Lee} CU, {Lee} HM, {Kutyrev} A, {Lim} G,
  {S{\'a}nchez-Ram{\'\i}rez} R, {Veilleux} S, {Wieringa} MH, {Yoon} Y (2017)
  {The X-ray counterpart to the gravitational-wave event GW170817}. Nature
  551(7678):71--74, \doi{10.1038/nature24290}, \eprint{1710.05433}

\bibitem[{{Ubertini} et~al.(2003){Ubertini}, {Lebrun}, {Di Cocco}, {Bazzano},
  {Bird}, {Broenstad}, {Goldwurm}, {La Rosa}, {Labanti}, {Laurent}, {Mirabel},
  {Quadrini}, {Ramsey}, {Reglero}, {Sabau}, {Sacco}, {Staubert}, {Vigroux},
  {Weisskopf}, and {Zdziarski}}]{Ubertini2003}
{Ubertini} P, {Lebrun} F, {Di Cocco} G, {Bazzano} A, {Bird} AJ, {Broenstad} K,
  {Goldwurm} A, {La Rosa} G, {Labanti} C, {Laurent} P, {Mirabel} IF, {Quadrini}
  EM, {Ramsey} B, {Reglero} V, {Sabau} L, {Sacco} B, {Staubert} R, {Vigroux} L,
  {Weisskopf} MC, {Zdziarski} AA (2003) {IBIS: The Imager on-board INTEGRAL}.
  Astronomy and Astrophysics 411:L131--L139, \doi{10.1051/0004-6361:20031224}

\bibitem[{{van Paradijs} et~al.(1997){van Paradijs}, {Groot}, {Galama},
  {Kouveliotou}, {Strom}, {Telting}, {Rutten}, {Fishman}, {Meegan}, {Pettini},
  {Tanvir}, {Bloom}, {Pedersen}, {N{\o}rdgaard-Nielsen}, {Linden-V{\o}rnle},
  {Melnick}, {Van der Steene}, {Bremer}, {Naber}, {Heise}, {in't Zand},
  {Costa}, {Feroci}, {Piro}, {Frontera}, {Zavattini}, {Nicastro}, {Palazzi},
  {Bennett}, {Hanlon}, and {Parmar}}]{VanParadijs97}
{van Paradijs} J, {Groot} PJ, {Galama} T, {Kouveliotou} C, {Strom} RG,
  {Telting} J, {Rutten} RGM, {Fishman} GJ, {Meegan} CA, {Pettini} M, {Tanvir}
  N, {Bloom} J, {Pedersen} H, {N{\o}rdgaard-Nielsen} HU, {Linden-V{\o}rnle} M,
  {Melnick} J, {Van der Steene} G, {Bremer} M, {Naber} R, {Heise} J, {in't
  Zand} J, {Costa} E, {Feroci} M, {Piro} L, {Frontera} F, {Zavattini} G,
  {Nicastro} L, {Palazzi} E, {Bennett} K, {Hanlon} L, {Parmar} A (1997)
  {Transient optical emission from the error box of the
  {\ensuremath{\gamma}}-ray burst of 28 February 1997}. Nature
  386(6626):686--689, \doi{10.1038/386686a0}

\bibitem[{{Vigneron} et~al.(2018){Vigneron}, {Lodato}, and
  {Guidarelli}}]{Vigneron18}
{Vigneron} Q, {Lodato} G, {Guidarelli} A (2018) {Tidal disruption of stars in a
  supermassive black hole binary system: the influence of orbital properties on
  fallback and accretion rates}. Monthly Notices of the RAS 476(4):5312--5322,
  \doi{10.1093/mnras/sty585}, \eprint{1803.05009}

\bibitem[{{von Borzeszkowski} and {Treder}(1988)}]{Borzes88}
{von Borzeszkowski} HH, {Treder} HJ (1988) {The meaning of quantum gravity},
  vol~20. Springer Netherlands, \doi{10.1007/978-94-009-3893-9}

\bibitem[{{Walker} et~al.(2000){Walker}, {Schaefer}, and {Fenimore}}]{Walker00}
{Walker} KC, {Schaefer} BE, {Fenimore} EE (2000) {Gamma-Ray Bursts Have
  Millisecond Variability}. Astrophysical Journal 537(1):264--269,
  \doi{10.1086/308995}

\bibitem[{{Wei} and {Wu}(2017)}]{Wei17}
{Wei} JJ, {Wu} XF (2017) {A Further Test of Lorentz Violation from the
  Rest-frame Spectral Lags of Gamma-Ray Bursts}. Astrophysical Journal
  851(2):127, \doi{10.3847/1538-4357/aa9d8d}, \eprint{1711.09185}

\bibitem[{{Xu} and {Ma}(2016{\natexlab{a}})}]{Xu2016b}
{Xu} H, {Ma} BQ (2016{\natexlab{a}}) {Light speed variation from gamma ray
  burst GRB 160509A}. Physics Letters B 760:602--604,
  \doi{10.1016/j.physletb.2016.07.044}, \eprint{1607.08043}

\bibitem[{{Xu} and {Ma}(2016{\natexlab{b}})}]{Xu2016a}
{Xu} H, {Ma} BQ (2016{\natexlab{b}}) {Light speed variation from gamma-ray
  bursts}. Astroparticle Physics 82:72--76,
  \doi{10.1016/j.astropartphys.2016.05.008}, \eprint{1607.03203}

\bibitem[{Yoneya(1987)}]{Yoneya87}
Yoneya T (1987) Wandering in the fields. World Scientific

\bibitem[{Yoneya(1989)}]{Yoneya89}
Yoneya T (1989) {On the Interpretation of Minimal Length in String Theories}.
  Mod Phys Lett A 4:1587, \doi{10.1142/S0217732389001817}

\bibitem[{Yoneya(1997)}]{Yoneya97}
Yoneya T (1997) D-particles, d-instantons, and a space-time uncertainty
  principle in string theory. \eprint{hep-th/9707002}

\bibitem[{{Zhang} et~al.(2019){Zhang}, {Kole}, {Bao}, {Batsch}, {Bernasconi},
  {Cadoux}, {Chai}, {Dai}, {Dong}, {Gauvin}, {Hajdas}, {Lan}, {Li}, {Li}, {Li},
  {Liu}, {Liu}, {Marcinkowski}, {Produit}, {Orsi}, {Pohl}, {Rybka}, {Shi},
  {Song}, {Sun}, {Szabelski}, {Tymieniecka}, {Wang}, {Wang}, {Wen}, {Wu}, {Wu},
  {Wu}, {Xiao}, {Xiong}, {Zhang}, {Zhang}, {Zhang}, {Zhang}, and
  {Zwolinska}}]{Zhang2019}
{Zhang} SN, {Kole} M, {Bao} TW, {Batsch} T, {Bernasconi} T, {Cadoux} F, {Chai}
  JY, {Dai} ZG, {Dong} YW, {Gauvin} N, {Hajdas} W, {Lan} MX, {Li} HC, {Li} L,
  {Li} ZH, {Liu} JT, {Liu} X, {Marcinkowski} R, {Produit} N, {Orsi} S, {Pohl}
  M, {Rybka} D, {Shi} HL, {Song} LM, {Sun} JC, {Szabelski} J, {Tymieniecka} T,
  {Wang} RJ, {Wang} YH, {Wen} X, {Wu} BB, {Wu} X, {Wu} XF, {Xiao} HL, {Xiong}
  SL, {Zhang} LY, {Zhang} L, {Zhang} XF, {Zhang} YJ, {Zwolinska} A (2019)
  {Detailed polarization measurements of the prompt emission of five gamma-ray
  bursts}. Nature Astronomy 3:258--264, \doi{10.1038/s41550-018-0664-0},
  \eprint{1901.04207}

\end{thebibliography}

\end{document}